\documentclass[singlecolumn]{article}

\usepackage{arxiv}

\usepackage[utf8]{inputenc} 
\usepackage[T1]{fontenc}    
\usepackage{hyperref}       
\usepackage{url}            
\usepackage{booktabs}       
\usepackage{amsfonts}       
\usepackage{nicefrac}       
\usepackage{microtype}      
\usepackage{graphicx}
\usepackage[sort]{natbib}
\setcitestyle{comma,numbers,open={[},close={]}}
\usepackage{doi}

\title{A dual-stage constitutive modeling framework based on finite strain data-driven identification and physics-augmented neural networks}

\author{Lennart Linden\thanks{Lennart Linden and Karl A. Kalina have contributed equally to this work.}\\
	Chair of Computational and\\
	Experimental Solid Mechanics\\
	TU Dresden,
	01062 Dresden, Germany \\
	\And
	Karl A. Kalina${}^*$\\
	Chair of Computational and\\
	Experimental Solid Mechanics\\
	TU Dresden,
	01062 Dresden, Germany \\
	\And
	J\"{o}rg Brummund\\
	Chair of Computational and\\
	Experimental Solid Mechanics\\
	TU Dresden,
	01062 Dresden, Germany \\
	\And
	Brain Riemer\\
	Chair of Computational and\\
	Experimental Solid Mechanics\\
	TU Dresden,
	01062 Dresden, Germany \\
	\And
	Markus K\"{a}stner\thanks{Corresponding author, email: \texttt{markus.kaestner@tu-dresden.de}.} \\
	Chair of Computational and\\
	Experimental Solid Mechanics\\
	TU Dresden, 
	01062 Dresden, Germany \\
}

\renewcommand{\shorttitle}{Dual-stage constitutive modeling framework}

\hypersetup{
pdftitle={Preprint_LindenEtAl_2025},
pdfsubject={},
pdfauthor={Linden},
pdfkeywords={automated constitutive modeling, data-driven identification, physics-augmented neural networks, finite strain, hyperelasticity},
}

\usepackage{pifont}

\usepackage{amssymb}

\usepackage{amsthm}
\usepackage{amsmath}
\usepackage{siunitx}  
\usepackage{mathtools}		
\mathtoolsset{centercolon}	
\usepackage{newtxmath}      
\usepackage{lineno}
\usepackage{bbm}
\usepackage[dvipsnames]{xcolor}
\usepackage{multirow}
\usepackage{upgreek}

\usepackage{epstopdf} 
\usepackage[normalem]{ulem}

\newcommand{\Softplus}{\mathscr{S\! P}}

\newcommand{\tr}{\operatorname{tr}}

\usepackage{enumitem}
\usepackage{tikz}
\newcommand*\circled[1]{\tikz[baseline=(char.base)]{
            \node[shape=circle,draw,inner sep=2pt] (char) {#1};}}

\newcommand\smallcircled[1]{\tikz[baseline=(char.base)]{
            \node[shape=circle, draw, inner sep=1pt] (char) {#1};}}

\newcommand{\Sym}{\mathscr{S\! y\! m}} 
\newcommand{\Orth}{\mathscr{O}}
\newcommand{\SO}{\mathscr{S\!O}}
\newcommand{\GL}{\mathscr{G\!\!L}}

\newcommand{\ve}[1]{\boldsymbol{#1}}
\newcommand{\te}[1]{\mathbf{#1}}
\newcommand{\teg}[1]{\boldsymbol{#1}}

\newcommand{\diffp}[2]{\frac{\partial #1}{\partial #2}}

\newcommand{\V}[1]{\underline{\mathbf{#1}}}
\newcommand{\Vg}[1]{\underline{\boldsymbol{#1}}}
\newcommand{\M}[1]{\uuline{\boldsymbol #1}}

\DeclareMathOperator{\trace}{tr}

\DeclareMathOperator{\Cof}{cof}

\DeclareMathOperator{\rank}{rank}
\DeclareMathOperator{\sym}{sym}
\newcommand{\cof}{\Cof}

\newcommand{\nablaX}{\nabla_{\!\!{\ve X}}}

\newcommand{\tte}[1]{\mathbb{#1}} 
\newcommand{\ttes}[1]{\mathbbm{#1}} 

\newcommand{\numbers}[1]{\mathbb{#1}}
\newcommand{\N}{\numbers{N}}    
\newcommand{\R}{\numbers{R}}    
\newcommand{\Lspace}[1]{\mathcal{L}_{#1}}

\usepackage{stackengine}
\usepackage{scalerel}

\usepackage[ruled,vlined]{algorithm2e}
\SetKw{KwAll}{all}


\theoremstyle{definition} 
\newtheorem{rmk}{Remark}
\usepackage[font=small]{caption}

\begin{document}

\maketitle

\begin{abstract}
In this contribution, we present a novel consistent dual-stage approach for the automated generation of hyperelastic constitutive models which only requires experimentally measurable data. To generate input data for our approach, an experiment with full-field measurement has to be conducted to gather testing force and corresponding displacement field of the sample. 
Then, in the first step of the dual-stage framework, a new finite strain Data-Driven Identification (DDI) formulation is applied. This method enables to identify tuples consisting of stresses and strains by only prescribing the applied boundary conditions and the measured displacement field. In the second step, the data set is used to calibrate a Physics-Augmented Neural Network (PANN), which fulfills all common conditions of hyperelasticity by construction and is very flexible at the same time.
We demonstrate the applicability of our approach by several descriptive examples. Two-dimensional synthetic data are exemplarily generated in virtual experiments by using a reference constitutive model. The calibrated PANN is then applied in 3D Finite Element simulations. In addition, a real experiment including noisy data is mimicked.
\end{abstract}

\section*{Graphical abstract}
\begin{center}
\includegraphics[width=0.8\textwidth]{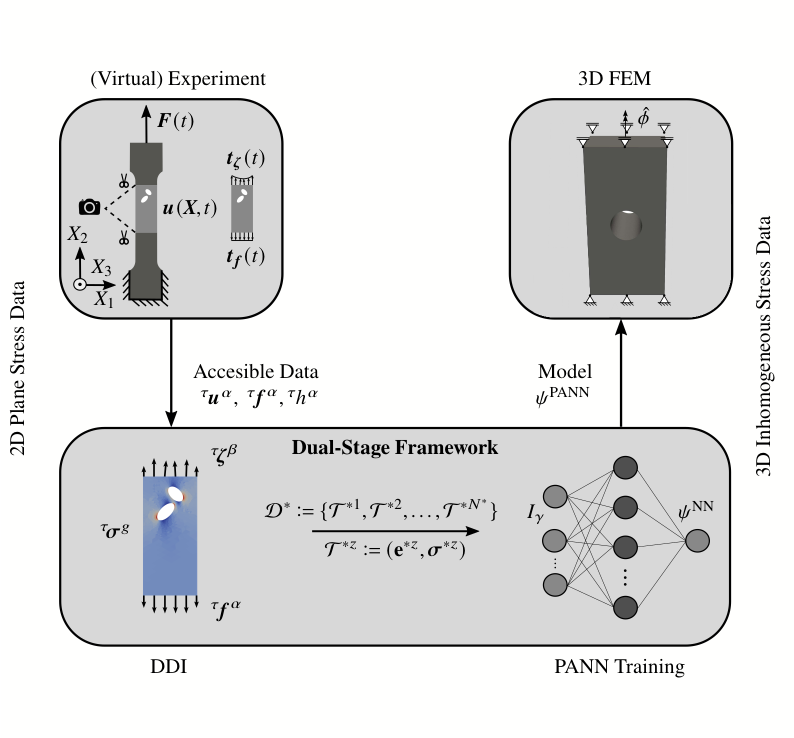}
\end{center}

\keywords{automated constitutive modeling \and data-driven identification \and physics-augmented neural networks \and \\ finite strain \and hyperelasticity}	
	
	\section{Introduction}
	\label{sec:intro}
	One of the fundamental elements of continuum mechanics are constitutive models which mathematically describe the behavior of materials. During the last century, considerable efforts have been made to understand the mathematical and physical requirements that a constitutive model should fulfill \cite{Haupt2000,Holzapfel2000}. Building on this understanding, plenty of classical constitutive models have been developed and parameterized using experimental data or simulation data from lower scales. Alternatives based on \emph{machine learning} and in particular the use of \emph{Neural Networks (NNs)} are often more flexible and can be discovered automatically \cite{Fuhg2024c}. 
	
	\subsection{Modeling of elasticity with neural networks}
	In their groundbreaking work from the early 1990s, Ghaboussi~et~al.~\cite{Ghaboussi1991} were the first to utilize NNs, specifically \emph{Feedforward Neural Networks (FNNs)}, to predict hysteresis under both uniaxial and multiaxial stress conditions. 
	After a brief surge in the 1990s, the use of NNs for constitutive modeling was largely set aside for some time. However, with the rise of machine learning and rapid advancements in its efficiency and accessibility, numerous NN-based constitutive models have emerged in recent years within a short period.
	This resurgence has been primarily driven by the growing trend of incorporating fundamental physics into NN-based constitutive models and scientific machine learning frameworks in general. This approach is often referred to as \emph{physics-informed} \cite{Raissi2019,Anton2023,Bastek2023,Harandi2024}, \emph{mechanics-informed} \cite{Asad2022}, \emph{physics-augmented} \cite{Klein2022,Linden2023}, \emph{physics-constrained} \cite{Kalina2023}, or \emph{thermodynamics-based} \cite{Masi2021}. Physical constraints can be integrated in two main ways: strongly, by modifying the network architecture \cite{Kalina2022a,Linka2021}, or weakly, through the use of problem-specific loss functions during training \cite{Rosenkranz2023,Weber2021,Weber2023}. As demonstrated in studies such as \cite{Linden2023,Masi2021,Fuhg2023,Rosenkranz2024,Masi2024}, these approaches enable the use of sparse training data while significantly enhancing the models' ability to extrapolate.
	
	Numerous groups have employed NNs to model elasticity in this context. For instance, the early works from the 2000s \cite{Shen2004,Liang2008} approximated the elastic potential of isotropic materials using FNNs, where the input consisted of three deformation-type invariants. These approaches inherently satisfied several requirements, such as thermodynamic consistency, objectivity, and material symmetry but were trained directly using strain energy density data which is usually not accessible.
	In recent years, architectures that output the hyperelastic potential and use invariants as inputs have become widespread \cite{Linden2023,Linden2021,Klein2021,Kalina2022a,Thakolkaran2022,Linka2021,Fuhg2022b,Tac2024a,Bahmani2024,Benady2024,Peirlinck2024}. A key aspect of these approaches is the loss function, which involves computing the derivative of the energy with respect to the deformation, a method known as \emph{Sobolev training} \cite{Czarnecki2017,Vlassis2020,Vlassis2021,Kalina2024}. 
	
	Moreover, several approaches incorporate polyconvex NNs \cite{Klein2021,Tac2022a,Chen2022,Linka2023,Tac2024a,Bahmani2024,Dammass2025a,Dammass2025}, which are particularly beneficial for \emph{Finite Element (FE)} simulations. These models improve extrapolation capabilities \cite{Linden2023,Kalina2024} and ensure rank-one convexity and thus ellipticity \cite{Ebbing2010,Schroder2010}. A popular approach for achieving this condition is the use of \emph{Fully Input Convex Neural Networks (FICNNs)}, introduced by Amos~et~al.~\cite{Amos2017}. 	
	Recently, Linden~et~al.~\cite{Linden2023} proposed a method based on invariants and FICNNs that inherently satisfies all standard conditions for (an)isotropic compressible hyperelasticity. This concept is denoted as \emph{Physics-Augmented Neural Network (PANN)}. An alternative approach is to use principal stretches combined with FICNNs as suggested by Vijayakumaran~et~al.~\cite{Vijayakumaran2025}.

	\subsection{Data mining and model identification from full field measurements}
	To represent the behavior of real materials with a constitutive model, \emph{experimental data} is required. Thereby, in general, two primary sources for data mining can be distinguished: (i) common experimental setups as uniaxial or biaxial tensile tests which result in homogeneous strain states, and (ii) experimental setups involving specimens that lead to inhomogeneous strain fields under loading, combined with full-field displacement measurements.\footnote{A third source of data are lower-scale simulations that enable to determine the homogenized response of a material. However, in order to apply this technique, knowledge about the material's microstructure and the constitutive behavior of the components, e.g., fibers and matrix, is required.}
	Experiments that belong to group (i) are less complex to perform, but only allow the recording of a limited range of deformation and stress states.
	In contrast, the experiments of group (ii) allow the recording of an extensive database with a single test. Although this advantage is accompanied by a more complex measurement setup and a more demanding post-evaluation, group (ii) is advantageous for data mining in the context of data-driven approaches, which often require more data compared to classical models. In the following, we will give a brief overview on techniques to evaluate full-field measurements.
		
	In contrast to the methods belonging to (i), the direct determination of stress-strain data is not possible in group (ii). Although inhomogeneous displacement and strain fields can be determined by means of \emph{Digital Image Correlation (DIC)} in the 2D case \cite{Pierron2020} and by using \emph{in-situ Computer Tomography (CT)} also in 3D case \cite{Lenoir2007}, the direct measurement of stresses is not possible. There are basically two ways to tackle the problem: (ii.a) the experiment is replicated within a simulation including a constitutive model and the model parameters are determined by solving the \emph{inverse problem}, and (ii.b) the experiment is reproduced in a \emph{model-free data-driven approach} without choosing a specific constitutive model and a data set consisting of stresses and strains or time sequences of these quantities is determined instead of model parameters.
	
	In group (ii.a), there are numerous approaches that have been developed over several decades, see the overviews in Avril~et~al.~\cite{Avril2008}, Roux~and~Hild~\cite{Roux2020} or Römer~et~al.~\cite{Romer2024}. In the works \cite{Mahnken1996,Kleuter2007}, for example, the inverse determination of the model parameters of a classical model is carried out by iteratively performing FE simulations reproducing the experiment, whereby the deviation between simulation and experiment is minimized by updating the model parameters. This technique is often called \emph{Finite Element Model Updating (FEMU)} method \cite{Avril2008,Romer2024}. In addition, numerous approaches specially adapted for inverse problems have also been developed, e.g., \emph{Virtual Fields Method (VFM)} \cite{Pierron2012} or \emph{modified Constitutive Relation Error (mCRE)}.
	In recent years, the described techniques have been applied in a data-driven context or novel data-driven approaches have been developed. For example, the \emph{EUCLID (Efficient Unsupervised Constitutive Law Identification and Discovery)} framework is an extension of the VFM. Both parameters and suitable constitutive models from a model catalog are determined by sparse regression \cite{Flaschel2021,Flaschel2023}. In \emph{NN-EUCLID} \cite{Thakolkaran2022}, the model catalog is replaced with an NN-based constitutive model and weights and biases are determined, see also \cite{Meng2025} for an application to 3D displacement fields obtained in a bulge inflation.
	In the same line, \cite{Lourenco2024} explores the integration of \emph{Recurrent Neural Network (RNN)}-based constitutive models into the VFM.
	The \emph{NN-mCRE (Neural Network modified Constitutive Relation Error)} approach builds up on the mCRE concept and replaces the classical constitutive model with a PANN \cite{Benady2024,Benady2024a}. Finally, in \cite{Wu2025} PANNs are combined with FEMU and in \cite{Wiesheier2024} a constitutive model based on splines is formulated and calibrated via FEMU. In the works \cite{Anton2023,Hamel2023}, \emph{Physics-Informed Neural Networks (PINNs)} are used to solve inverse problems and thus calibrating constitutive models.

	The comparatively new approaches belonging to group (ii.b) enable to generate stress-strain data from experiments with full-field measurement without selecting a specific constitutive model for solving the inverse problem. The most widespread method in this group is the \emph{Data-Driven Identification (DDI)} introduced by Leygue~et~al.~\cite{Leygue2018,Leygue2019} and Stainier~et~al.~\cite{Stainier2019}. This method is based on the \emph{Data-Driven Computational Mechanics (DDCM)} approach by Kirchdoerfer~and~Ortiz~\cite{Kirchdoerfer2016}.\footnote{DDCM completely avoids to use constitutive equations. Instead, a data set consisting of stress-strain tuples characterizing the material’s behavior is used. Thus, a data-driven solver seeks to minimize the distance between the searched solution and the material data set within a proper energy norm, while compatibility and equilibrium have to be satisfied simultaneously \cite{Kirchdoerfer2016}. 
    Several extensions of the DDCM have been proposed, e.g., for inelasticity \cite{Eggersmann2019,Ciftci2022,Bartel2023}, fracture mechanics \cite{Carrara2020} or multiscale problems \cite{Karapiperis2021}.
    Comparisons of the DDCM method with NN-based constitutive modeling approaches can be found in the works \cite{Stocker2024,Zlatic2024}.} An application of DDI to real experiments of thin perforated disks as well as an extension to finite deformations is shown in \cite{Dalemat2019}. The robustness against incomplete input data is discussed in \cite{Dalemat2024}. An extension of DDI enabling the consideration of uncertain forces is introduced in \cite{Zschocke2023}. Furthermore, locally convex reconstruction methods are integrated into DDI in \cite{Su2022}, which improves accuracy and efficiency compared to the original approach.
	Finally, the modified DDI approach presented in \cite{Langlois2022} can even be applied to certain classes of inelastic materials.
	Another alternative for determining heterogeneous stress fields without a specific constitutive model are the approaches \cite{Cameron2021a,Liu2021a,Cameron2022,Cameron2023}, in which the momentum balance is solved numerically. Like DDI, these approaches require a displacement field from which strains can be determined, as well as specified stresses at the boundary. However, the latter methods are limited to isotropic materials.
	
	\subsection{Objectives and contributions of this work}
	
	As can be seen from the literature review, in the vast majority of cases, the application of many approaches presupposes that at least the basic material class is known in advance, e.g., elasticity or plasticity. This can be avoided by using model-free approaches to data generation, e.g., DDI.
	
	We present a consistent \emph{data-driven dual-stage approach} for the automated constitutive modeling at finite strains, which only requires experimentally measurable data. As a pre-processing step, a (virtual) experiment with full-field measurement is required to generate raw input data. 
    Then, in the first step of our dual-stage approach, the recorded data is used as input for DDI to determine a data set of stress and strain tuples. The original DDI formulation \cite{Leygue2018} is extended to a new total Lagrangian finite strain formulation.
    In the second step of the proposed framework, the data set generated by the DDI is used to calibrate a hyperelastic PANN model \cite{Linden2023}. The calibrated PANN can then be used in 3D FE simulations.
	We demonstrate the applicability of our approach by several descriptive examples. Therefore, two-dimensional synthetic data are exemplarily generated by using a reference constitutive model. The calibrated PANN is then applied in three-dimensional FE simulations, where the solution is compared to the reference model. In addition, the behavior of our framework for noisy data is analyzed and the conditions of a real experiment are mimicked.
	
	The organization of the paper is as follows: In Sect.~\ref{sec:basics}, the fundamentals of finite strain continuum mechanics and basic principles of hyperelasticity are summarized. After this, the dual-stage framework is introduced in Sect.~\ref{sec:dual_stage}. This is followed by an introduction of the DDI in Sect.~\ref{sec:data_driven_identification} and the PANN model in Sect.~\ref{sec:neural_networks}. The developed approach is exemplarily applied in Sect.~\ref{sec:examples} within several examples. After a discussion of the results, the paper is closed by concluding remarks and an outlook to necessary future work in Sect.~\ref{sec:conclusion}.

	\paragraph{Notation}

    Throughout this work, the space of tensors $\Lspace{n}$ in the three-dimensional Euclidean vector space $\R^3$ is used, where $\N$ denotes the set of natural numbers without zero.
    Tensors of rank one and two are given by symbols in bold italics and upright print respectively, i.\,e., $\ve a \in \Lspace{1}$ or $\te b,\te c \in \Lspace{2}$. 
    Transpose and inverse of a second order tensor $\te b$ are marked by $\te b^\text{T}$ and $\te b^{-1}$, respectively. Furthermore, trace, determinant and cofactor are denoted by $\trace \te b$, $\det \te b$ and $\cof \te b := \det(\te b) \te b^\text{-T}$.
    The set of invertible second order tensors with positive determinant is denoted by $\GL^+(3):=\left\{\te A \in \Lspace{2}\,|\,\det \te A > 0\right\}$, while the orthogonal group and special orthogonal group in $\R^3$ are denoted by $\Orth(3):=\left\{\te A \in \Lspace{2}\,|\,\te A^\text{T} \cdot \te A = \te 1\right\}$ and $\SO(3):=\left\{\te A \in \Lspace{2}\,|\,\te A^\text{T} \cdot \te A = \te 1,\,\det \te A = 1\right\}$, respectively. Here, $\te 1\in \Lspace{2}$ denotes the second order identity tensor. 
    The space of symmetric second order tensors is denoted as \mbox{$\Sym:=\left\{\te A \in \Lspace{2} \, |\, \te A = \te A^\text{T}\right\}$}.
    Furthermore, the single and double contraction of two tensors are given by $\te b \cdot \te c = b_{kq} c_{ql} \ve e_k \otimes \ve e_l$ and $\te b:\te c=b_{kl}c_{lk}$, respectively. Thereby, $\ve e_k\in \Lspace{1}$ denotes a Cartesian basis vector and $\otimes$ is the dyadic product. The Einstein summation convention is used.
    $\nablaX$ is the nabla operator with respect to reference configuration $\mathcal{B}_0$.
    Moreover, fourth order tensors are denoted by $\tte A \in \Lspace{4}$ and the fourth order identy tensor with major and minor symmetry is given by $\mathbbm 1 =\frac{1}{2}(\delta_{km}\delta_{ln}+\delta_{kn}\delta_{lm})\ve e_k \otimes\ve e_l \otimes\ve e_m \otimes\ve e_n\in \Lspace{4}$.
    The inverse of a fourth order tensor $\tte A \in \Lspace{4}$ with major and minor symmetry is defined by satisfying the relation $\tte A : \tte A^{-1} = \mathbbm 1$.
    For reasons of readability, the arguments of functions are usually omitted within this work. However, to show the dependencies, energy functions are given with their arguments, except when derivatives are written. Also, we let the symbol of a function be identical to the symbol of the function value itself.

    \section{Fundamentals of hyperelasticity}\label{sec:basics}
    In this section, we briefly outline the fundamental kinematic and stress quantities along with general relations of isotropic hyperelastic constitutive models. For a more comprehensive overview, we refer to textbooks such as those by Haupt~\cite{Haupt2000} or Holzapfel~\cite{Holzapfel2000}.
    
    \subsection{Kinematics and stress measures}\label{sec:kinematic_stress}
    We consider a material body $\mathcal{K}$ occupying the reference configuration $\mathcal B_0 \subset \R^3$ at time $t_0 \in \R_{\geq 0}$ and the current configuration $\mathcal B \subset \R^3$ at time $t \in \mathcal{T} := \{\tau \in \R_{> 0} \, | \, \tau \ge t_0\}$. The displacement vector $\ve{u} \in \Lspace{1}$ of a material point $P \in \mathcal{K}$, relating the reference position $\ve{X} \in \mathcal B_0$ at $t_0$ and the current position $\ve{x} \in \mathcal B$ at $t$, is defined as $\ve{u}(\ve{X}, t) := \ve{\varphi}(\ve{X}, t) - \ve{X}$. Here, $\ve{\varphi}: \mathcal B_0 \times \mathcal{T} \to \mathcal B$, $(\ve{X}, t) \mapsto \ve{x} := \ve{\varphi}(\ve{X}, t)$ represents a bijective motion function, continuous in space and time. The deformation gradient $\te{F} \in \GL^+(3)$ and the Jacobian determinant $J \in \R_{> 0}$ are defined by
    \begin{align}
        \te{F} := (\nablaX \ve{\varphi})^\text{T} \quad \text{and} \quad J := \det \te{F} > 0 \; ,
    \end{align}
    where $\nablaX$ denotes the nabla operator with respect to the reference configuration $\mathcal B_0$.
    Deformation measures independent of rigid body motions are given by the right Cauchy-Green deformation tensor $\te{C} := \te{F}^\text{T} \cdot \te{F} \in \Sym$ and the left Cauchy-Green deformation tensor $\te{b} := \te{F} \cdot \te{F}^\text{T} \in \Sym$, which are positive definite, i.e., all eigenvalues are positive. In addition, we introduce the strain measures $\te E := \frac{1}{2}(\te C -\te 1)$ and $\te e:=\frac{1}{2}(\te 1 - \te b^{-1})$ denoted as Green-Lagrange and Euler-Almansi strain tensor, respectively.
    
    Within nonlinear continuum solid mechanics, various stress measures are defined. The symmetric Cauchy stress $\teg{\sigma} \in \Sym$, also known as true stress, refers to the current configuration $\mathcal B$. The Kirchhoff stress $\teg \tau \in \Sym$ and the 1st and 2nd Piola-Kirchhoff stress tensors $\te{P} \in \mathcal{L}_2$ and $\te{T} \in \Sym$ are obtained through pull-back operations: $\teg \tau := J \teg \sigma$, 
    $\te{P} := J \teg{\sigma} \cdot \te{F}^{-T}$ and $\te{T} := J \te{F}^{-1} \cdot \teg{\sigma} \cdot \te{F}^{-\text{T}}$.
    The stress tensors are related to the corresponding stress vectors $\ve t, \ve p, \ve T\in \Lspace{1}$ via Cauchy's theorem, i.e., $\ve t=\teg \sigma \cdot \ve n$, $J\ve t=\teg \tau \cdot \ve n$, $\ve p=\te P \cdot \ve N$ and $\ve T = \te T \cdot \ve N$, where $\ve n \in \Lspace{1}$ and $\ve N \in \Lspace{1}$ are unit normal vectors referred to $\mathcal B$ and $\mathcal B_0$, respectively.

    Following Haupt~\cite{Haupt2000}, the pairs $(\te e,\teg \tau)$ and $(\te E,\te T)$ build sets of \emph{conjugated or dual variables}, which is important for the introduced finite strain DDI formulations given in Sect.~\ref{sec:data_driven_identification}.

    \subsection{Hyperelasticity}
    \label{sec:Hyper}
    
    The constitutive behavior of the considered solids is restricted to \emph{hyperelasticity} within this work. Accordingly, a hyperelastic potential that is equal to the Helmholtz free energy density function \(\psi: \GL^+(3) \to \R_{\geq 0}, \ \te{F} \mapsto \psi(\te{F})\) exists. Thus, the \emph{thermodynamic consistency} of any hyperelastic model is fulfilled a priori by using the relation
    \begin{align}
        \label{eq:consistent}
        \te{P} = \diffp{\psi}{\te{F}}\; ,
    \end{align}
    which follows from the evaluation of the Clausius-Duhem inequality \cite{Holzapfel2000}.
    In addition, there are several further requirements on $\psi(\te F)$ that ensure a physically reasonable constitutive behavior \cite{Holzapfel2000,Linden2023}:
    
    \begin{itemize}
        \item \emph{Energy normalization condition:} The free energy density function must satisfy \(\psi(\te{1}) = 0\), meaning that the free energy vanishes in the undeformed state.
        
        \item \emph{Non-negativity condition:} The free energy should be non-negative for all admissible deformations \(\te{F} \in \GL^+(3)\), expressed as \(\psi(\te{F}) \geq 0 \).
    
        \item \emph{Stress normalization condition:} The undeformed state $\te F = \te 1$ should be stress-free, i.e., \(\te P = \te 0\).
    
        \item \emph{Volumetric growth condition:} This condition necessitates that an infinite amount of energy is required to infinitely expand the volume or compress it to zero, i.e., \(\psi(\te{F}) \to \infty\) must hold as \(J \to \infty\) or \(J \to 0^+\).
    
        \item \emph{Material objectivity:} The free energy must be invariant with respect to superimposed rigid body motions, expressed as \(\psi(\te{F}) = \psi(\te{Q} \cdot \te{F})\) for all special orthogonal tensors \(\te{Q} \in \SO(3)\). 
        Note that in hyperelasticity the fulfilment of objectivity automatically implies the required symmetry in $\teg \sigma \in \Sym$ and $\te T\in \Sym$ and thus the compatibility to the balance of angular momentum, cf. {\v S}ilhav{\'y}~\cite[Prop.~8.3.2]{Silhavy1997}.
                
        \item \emph{Material symmetry:} The constitutive equations should also reflect the material's underlying (an)isotropy which is expressed as material symmetry and formulated as \(\psi(\te F \cdot \te Q^\text{T}) = \psi( \te F)\quad\forall \,\te F \in \GL^+(3),\,\te Q \in \mathscr{G}\subseteq \Orth(3)\), where $\mathscr{G}$ denotes the symmetry group of the material under consideration.
    
        \item \emph{Polyconvexity:} Another condition that, unlike the other conditions, is not absolutely required is that the energy function \(\psi(\te F)\) fulfills polyconvexity, meaning it exists a representation $\psi(\te F) = \mathcal P(\te F, \cof \te F, \det \te F)$, where $\mathcal P(\te F, \cof \te F, \det \te F)$ is convex w.r.t. its arguments. This condition implies Legendre-Hadamard ellipticity, which ensures material stability, as detailed in Ebbing \cite{Ebbing2010}. However, polyconvexity is a relatively strong requirement on the free energy, see \cite{Kalina2024}.
    \end{itemize}
    To satisfy \emph{material objectivity}, which implies compatibility with the \emph{balance of angular momentum} \cite{Silhavy1997}, and \emph{material symmetry} condition automatically, the strain energy function is typically expressed in terms of invariants of the right Cauchy-Green deformation tensor and structural tensors capturing the underlying anisotropy \cite{Kalina2025,Jadoon2024}.
    Thus, for \emph{isotropic materials} with $\mathscr{G} = \Orth(3)$, the strain energy function is expressed in terms of invariants of the right Cauchy-Green deformation tensor $\te C$ which are equal to the invariants of the left Caucy-Green deformation tensor $\te b$, i.e., it holds \(\psi = \psi(I_1, I_2, I_3)\), with $I_1 = \tr\te{C} = \tr\te{b}$, $I_2 = \tr\left(\cof \te C\right) = \tr\left(\cof \te b\right)$ and $I_3 = \det\te{C} = \det\te{b}$. By using the chain rule and the pull-back/push-forward operations for the introduced stress measures, we find the following relations from Eq.~\eqref{eq:consistent}:
    \begin{align}
    \label{eq:stress_invariants}
        \te{T} = 2\sum_{\gamma=1}^3\diffp{\psi}{I_\gamma}\diffp{I_\gamma}{\te{C}} \quad \text{and} \quad \teg{\sigma} = \frac{2}{J} \sum_{\gamma=1}^3\diffp{\psi}{I_\gamma}\diffp{I_\gamma}{\te{b}} \cdot \te{b}\; 
        \, .
    \end{align}

    \section{Dual-stage framework for automated data-driven constitutive modeling}\label{sec:dual_stage}

    \subsection{Framework}\label{sec:framework}
\begin{figure}[!t]
    	    \centering
    	    \def\svgwidth{\linewidth}\small{
    	    \includegraphics[width=0.8\textwidth]{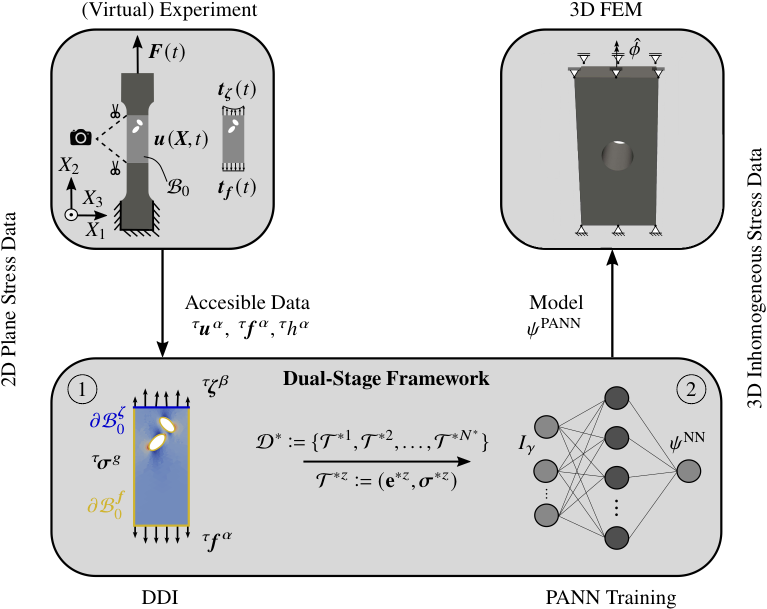}}
    	    \caption{Dual-stage framework: \protect\smallcircled{1} Using DDI in order to generate material database $\mathcal D^*$ of stress-strain states based only on measurable full-field data $\ve u(\ve X,t)$ and global testing force $\ve F(t)$ from a (virtual) experiment as well as boundary conditions. \protect\smallcircled{2} PANN serves as constitutive model trained on determined material database and can be used for prediction of unseen data.}
    	    \label{fig:dual_stage_framework}
    \end{figure}

    Based on the summarized continuum theory, the following section introduces the proposed \emph{data-driven dual-stage framework} enabling an \emph{automated data-driven modeling}. The framework requires \emph{input data} that have to be generated within a data mining step beforehand. More specifically, a (virtual) experiment with full-field displacement measurement is needed.
    The general procedure of our dual-stage framework is basically subdivided into two steps referred as
    \begin{enumerate}[label=\protect\circled{\arabic*}]
    \item applying \emph{Data-Driven Identification (DDI)} to generate a material dataset $\mathcal D^*$ consisting of material states, i.e., stress-strain tuples $\mathcal T^{*z}=(\te e^{*z},\teg \sigma^{*z})$ or $\mathcal T^{*z}=(\te E^{*z},\te T^{*z})$ with $z\in \{1,\dots,N^*\}$ and
    \item training of a \emph{Physics-Augmented Neural Network (PANN) model} based on the dataset $\mathcal D^*$ with $\vert\mathcal D^*\vert = N^*$.
    \end{enumerate}
    A schematic representation of the dual-stage framework and the necessary pre-processing steps for mining the raw data and the subsequent application of the generated constitutive model in an FE simulation can be seen in Fig.~\ref{fig:dual_stage_framework}. In the absence of a real experiment, we generate input data for the framework by using a virtual experiment within this work. The two  main ingredients of the dual-stage framework, i.e., DDI and PANNs, are described in detail in Sect.~\ref{sec:data_driven_identification} and Sect.~\ref{sec:neural_networks}, respectively. 
    
    \paragraph{Data mining as a pre-processing step}
    To apply, the dual-stage approach, input data for the DDI has to be captured within a (virtual) experiment with full-field measurement, i.e., stress or traction vector $\ve t_{\!\ve f}(\ve X,t)$ on the domain boundary $\partial\mathcal B^{\ve f}$ and the displacement field $\ve u(\ve X,t)$ within the complete domain $\mathcal B$ corresponding to the loading situation, cf.~Fig.~\ref{fig:dual_stage_framework}. Flat specimens, leading to \emph{plane stress} states in a good approximation \cite{Kalina2022a}, with ellipsoidal holes are used to cover a wide range of deformation states \cite{Kalina2022a,Flaschel2021}. The displacement field, or more precisely the displacements at a finite number of control points, can be determined using DIC, cf. \cite{Pierron2020,Dalemat2019,Wiesheier2024}. The domain $\mathcal B$ does not include the entire sample but merely represents a suitably selected section, cf. Fig.~\ref{fig:dual_stage_framework}, (virtual) experiment.

    \begin{rmk}\label{remark:compressible}
        Note that for the consideration of \emph{compressible materials} the determination of the in-plane displacements is not sufficient because the complete 3D deformation state is not accessible. The change in thickness must also be determined, i.e., the 3D displacement field must be measured from both sides of the specimen.
        This is due to the fact that the full deformation state and the corresponding stress tensor is needed for the training of the PANN in step \protect\smallcircled{2} of the dual-stage approach. Assuming \emph{plane stress} and homogeneous deformations through the thickness, the following holds for deformation gradient and 1st Piola-Kirchhoff stress tensor:
        \begin{align}
            [\te F] = \begin{bmatrix}
                F_{11} & F_{12} & 0 \\
                F_{21} & F_{22} & 0 \\
                0 & 0 & \lambda_3 \\
            \end{bmatrix}
            \quad \text{and} \quad
            [\te P] = \begin{bmatrix}
                P_{11} & P_{12} & 0 \\
                P_{21} & P_{22} & 0 \\
                0 & 0 & 0 \\
            \end{bmatrix} \, ,
        \end{align}
        equivalently for the strain measures $\te e$ and $\te E$ as well as the stress tensors $\teg \sigma$ and $\te T$.
        Thereby, the stretch field in thickness direction can be determined via $\lambda_3(X_1,X_2,t)=[u_3^+(X_1,X_2,t) - u_3^-(X_1,X_2,t) + h_0]/h_0 \in \R_{>0}$, where $u_3^+(X_1,X_2,t)\in \R$ and $u_3^-(X_1,X_2,t)\in \R$ are displacements on opposite sites and $h_0\in \R_{>0}$ is the sample's thickness in the undeformed state, respectively. Thus, in addition to the in-plane displacement field, the out-off-plane stretching field is also transferred as input for the dual-stage approach, i.e., step \protect\smallcircled{1}, the DDI.
    \end{rmk}

    In contrast to the displacement field, the traction vector $\ve t$ is not directly measurable in experimental setups. Only the global testing force $\ve F(t) \in \Lspace{1}$ is accessible from the testing machine's loading cell.
    If the considered cross section is sufficiently far from local disturbances such as holes or clamping effects, the stress state can be assumed to be homogeneous.\footnote{\label{foot:anisotropy}For anisotropic materials, the assumption of a homogeneous stress state may not hold due to directional dependencies in the material properties, which can result in non-uniform stress distributions even at distances far from geometric or loading discontinuities.} This justifies the approximation $\ve t_{\!\ve f}(t) = \ve F(t)/A \in \Lspace{1}$, where $A\in\R_{>0}$ is the actual cross sectional area. Such a situation is depicted in Fig.~\ref{fig:dual_stage_framework} for a cross section partly coinciding with the domain boundary $\partial \mathcal B^{\ve f}$.
    For the upper boundary this assumption is not valid. Here, we can only state that 
    \begin{align}
        \ve F(t) = \int\limits_{\partial \mathcal B^{\ve \zeta}} \ve t_{\ve \zeta}(\ve X,t) \, \mathrm dA \; .
    \end{align}
    For free boundaries, the traction vector is identical to zero.

    
    
    \paragraph{Data-driven identification}
    Step \protect\smallcircled{1} of the dual-stage approach is the DDI \cite{Leygue2018}, in which a material database $\mathcal D^*:=\{\mathcal T^{*1}, \mathcal T^{*2}, \dots, \mathcal T^{*N^{*}}\}$ consisting of $N^*$ material states, i.e., tuples $\mathcal T^{*z}:=(\te e^{*z}, \teg \sigma^{*z})$ or $\mathcal T^{*z}:=(\te E^{*z}, \te T^{*z})$ of strains and stresses, is generated as the output only from the variables measurable in the experiment, i.e., displacement field $\ve u(\ve X,t)$ in $\mathcal B$ and global testing force $\ve F(t)$, or variables derived from them, i.e., the traction vector $\ve t_{\!\ve f}(t)$ and the stretch in thickness direction $\lambda_3(X_1,X_2,t)$.

    The special characteristic of the DDI is that no specific constitutive model is required to determine the database $\mathcal D^*$. Only elasticity is assumed.
    To this end, a sophisticated optimization problem is formulated, whereby compatibility and equilibrium are required as constraints. The former is achieved explicitly via the element-wise determination of the strains from the interpolation with the shape functions. In contrast, the equilibrium requirement is incorporated using the Lagrange parameter method \cite{Leygue2018}. Formulations of the DDI based on different stress and strain measures are possible, cf. Sect.~\ref{sec:data_driven_identification}.

    
    \paragraph{Training of constitutive model}
    Within the final step \protect\smallcircled{2}, the data $\mathcal D^*$ is used to train a PANN which serves as a constitutive model automatically accounting for all common conditions of finite strain hyperelasticity. Although only plane stress data is used for calibrating the model, stresses for complex 3D deformation states can be predicted with high accuracy, which is due to the invariant-based fromulation of the PANN model, cf. Sect.~\ref{sec:neural_networks}. This allows the application of the trained model in 3D FE simulations \cite{Kalina2022a,Linden2023}, cf. Fig.~\ref{fig:dual_stage_framework}, application of the model.

    \subsection{Data-Driven Identification}\label{sec:data_driven_identification}
        
\begin{figure}
    	    \centering
    	    \includegraphics{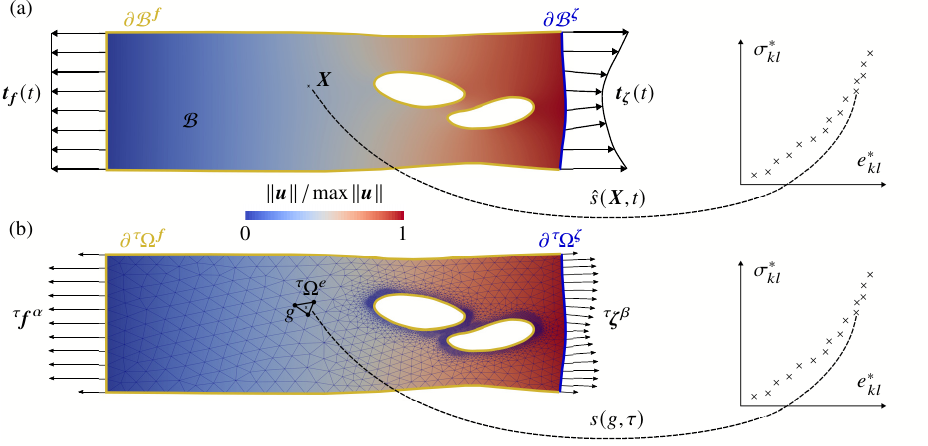}
    	    \caption{Starting point of DDI: (a) continuous specimen $\mathcal B$ under loading conditions with known stress vector $\ve{t_{\!\ve f}}(t)$ and unknown stress vector $\ve{t_{\ve\zeta}}(t)$, known displacement field $\ve u$ without rigid body translation and unknown material data base consisting of material states $(e_{kl}^{*z},\sigma_{kl}^{*z})$, which are associated to the material points $\ve X$ at time $t$ by the continuous mapping $\hat s(\ve X, t)$ as well as (b) specimen discretized into subdomains ${}^\tau\!\Omega^e$ under loading conditions with known nodal forces ${}^\tau\!\!\ve f^\alpha$ and unknown nodal forces ${}^\tau\!\ve \zeta^\beta$ at snapshot $\tau$ and unknown material data base consisting of material states $(e_{kl}^{*z},\sigma_{kl}^{*z})$, which are associated to the quadrature point $g$ of linear element $e$ at snapshot $\tau$ by the discrete mapping $s(g, \tau)$.
    	    }
    	    \label{fig:ddi_idea}
    	\end{figure}

    The method of Data-Driven Identification (DDI), introduced by Leygue~et~al.~\cite{Leygue2018,Leygue2019} and Stainier~et~al.~\cite{Stainier2019}, allows a set of admissible material states to be calculated without any assumption about the underlying constitutive behaviour, with the exception of elasticity.
    The motivation arises from a formulation based on continuous field quantities, as illustrated in Fig.~\ref{fig:ddi_idea}(a) and detailed in \ref{app:formulation_ddi}, which provides the transition to the discrete formulation.
    We will present three different finite strain DDI formulations in the following: An \emph{updated Lagrangian DDI formulation} based on the conjugated pair $(\te e, \teg \tau)$ and two \emph{total Lagrangian DDI formulations} both based on the conjugated pair $(\te E, \te T)$.

    \subsubsection{Updated Lagrangian formulation}\label{sec:ddi_updated_lagrange}
    
    The \emph{discrete loss term} in the \emph{updated Lagrangian DDI formulation}, as derived from the theoretical background in \ref{app:discrete_ddi}, is given by
    \begin{align}\label{eq:snapshots_loss_updated_lagrange}
    \begin{split}
        L^\text{UL} := &\frac{1}{2} \sum_\tau\sum_g w^g {}^\tau\!\!J^g_{\!\!\vartriangle} {}^\tau\!h^g \biggl[\left({}^\tau\!e_{kl}^g - e_{kl}^{*s(g,\tau)}\right) c_{klmn} \left({}^\tau\!e_{mn}^g - e_{mn}^{*s(g,\tau)}\right) 
        \\
        &+ \left({}^\tau\!\sigma_{ab}^g - \sigma_{ab}^{*s(g,\tau)}\right) c_{abcd}^{-1} \left({}^\tau\!\sigma_{cd}^g - \sigma_{cd}^{*s(g,\tau)}\right)\biggr] 
        \\
        &+  \sum_\tau\!\!\!\!\sum_{\substack{\alpha \\ \pi(\alpha) = 1}}\!\!\!\! {}^\tau\!\eta_c^\alpha \bigg({}^\tau\!\!f_c^\alpha - \sum_g w^g {}^\tau\!\!J^g_{\!\!\vartriangle} {}^\tau\!h^g {}^\tau\!b_{\!abc}^{\alpha g} {}^\tau\!\sigma_{ab}^g\bigg)
        \\
        &+  \sum_\tau\!\!\!\!\sum_{\substack{\beta \\ \pi(\beta) = 0}}\!\!\!\! {}^\tau\!\mu_c^\beta \bigg({}^\tau\!\zeta_c^\beta - \sum_g w^g {}^\tau\!\!J^g_{\!\!\vartriangle} {}^\tau\!h^g {}^\tau\!b_{\!abc}^{\beta g} {}^\tau\!\sigma_{ab}^g\bigg) \; ,
    \end{split}
    \end{align}
    where the function $\pi: \{1, \dots, N^\text{node}\} \to \{0, 1\}$ is defined as
    \begin{align}
    \label{eq:pi}
    \pi(\alpha) := \begin{cases} 
        1, & {}^\tau\!\!f_a^\alpha \, \text{is prescribed for all} \ \tau \in \{1, \dots, N^\text{snap}\} \ \text{in each direction} \ a \in \{1, 2\}\\
        0, & \text{else}
      \end{cases}
    \end{align}
    and serves as an indicator function.
    The pseudo stiffness tensor is chosen to $\ttes c = C \ttes 1$ with $C \in \R_{>0}$ \cite{Leygue2019}. The following notation is applied: Indices of the beginning of the Latin alphabet run from 1 to 2, e.g. $a,b,c,d\in\{1,2\}$. Indices from the middle of the Latin alphabet run from 1 to 3, e.g. $k,l,m,n\in\{1,2,3\}$.
    Einstein's summation convention is used for both types of indices. Small Greek superscript indices label the global node number and run from 1 to $N^\text{node}$, e.g., $\alpha\in\{1,\dots,N^\text{node}\}$. The top left superscript $\tau\in\{1,\dots,N^\text{snap}\}$ denotes the snapshot and the top right superscript $g\in\{1,\dots,N^\text{quad}\}$ denotes the quadrature point number. No Einstein summation convention is applied for all these three types of indices. Due to the plane stress assumption, all out-of-plane stress components are zero, i.e., ${}^\tau\!\sigma_{kl}^g = \sigma_{kl}^{*z} = 0$, if $k=3$ or $l=3$, respectively. For the strains, the out-of-plane shear components are also zero whereas ${}^\tau\!e_{33}^g\ne 0$ and $e_{33}^{*z}\ne 0$. Thereby, $z$ runs from 1 to $N^*$, i.e., $z \in \{1,\dots,N^*\}$.
    
    Fig.~\ref{fig:ddi_idea}(b) shows the basic idea of DDI in the discrete form: A loaded specimen meshed with triangles is analyzed within several snapshots $\tau$ for known nodal displacements ${}^\tau\!\ve u^\alpha$ and nodal forces ${}^\tau\!\ve f^\alpha$ to determine an unknown material database.\footnote{\label{foot:anisotropy_constraint}In the anisotropic case, it may be necessary to impose the additional condition that the global testing force equals the sum of nonzero nodal forces on the boundary, if the assumption of a homogeneous stress state does not hold, cf. Footnote~\ref{foot:anisotropy}. In this case, the optimization problem must be adjusted accordingly, as described in~\cite{Dalemat2019}.}
    The task at hand requires a customized staggered solution strategy, which was originally presented by Leygue~et~al.~\cite{Leygue2018} and is similar to the strategy developed for DDCM by Kirchdoerfer~and~Ortiz~\cite{Kirchdoerfer2016}. Below we present this scheme in a version tailored to our problem.

    \paragraph{Solution strategy}
    Within the DDI problem given by the loss term~\eqref{eq:snapshots_loss_updated_lagrange}, the following \emph{quantities are assumed to be known}, i.e., as described in Sect.~\ref{sec:framework}, measured within an experiment or determined from the measured quantities: 
    \begin{itemize}[noitemsep]
        \item reference nodal coordinates $(X_1^\alpha,X_2^\alpha)$ of a triangulated 2D domain, 
        \item in-plane nodal displacements ${}^\tau\! u_a^\alpha$ for all snapshots and nodes, 
        \item deformed thickness ${}^\tau\! h^g$ for all snapshots and quadrature points, 
        \item external nodal forces ${}^\tau\! f_a^\alpha$ for all nodes with $(X_1^\alpha, X_2^\alpha)\not\in \partial \Omega_0^{\ve \zeta}$ and for all snapshots, as well as
        \item strains ${}^\tau\! e_{kl}^g$ belonging to the mechanical states for all snapshots and quadrature points.
    \end{itemize}
    In addition, the number of material states $N^*$ and the pseudo stiffness tensor $\ttes c$ have to be prescribed. Conversely, the following \emph{variables are unknown} and must be calculated:
    \begin{itemize}[noitemsep]
        \item stresses ${}^\tau\! \sigma_{ab}^g$ belonging to the mechanical states for all snapshots and quadrature points,
        \item material states $e_{kl}^{*z}$ and $\sigma_{ab}^{*z}$, 
        \item discrete mapping $s(g,\tau)$ for all quadrature points and snapshots, 
        \item Lagrange multipliers ${}^\tau\! \eta_a^\alpha$ for all nodes with $(X_1^\alpha, X_2^\alpha)\not\in \partial \Omega_0^{\ve \zeta}$, 
        \item Lagrange multipliers ${}^\tau\! \mu_a^\alpha$ for all nodes with $(X_1^\alpha, X_2^\alpha)\in \partial \Omega_0^{\ve \zeta}$, as well as
        \item external nodal forces ${}^\tau\! \zeta_a^\alpha$ for all nodes with $(X_1^\alpha, X_2^\alpha)\in \partial \Omega_0^{\ve \zeta}$.
    \end{itemize}

    The goal is now to minimize the specified loss term $L^\text{UL}$. Thereby, the difficulty lies in the discrete mapping $s(g,\tau)\in\{1,\dots,N^*\}$, which excludes the use of typical gradient-based optimizers. We derive a staggered solution scheme according to \cite{Leygue2018}. To this end, we assume that the mapping $s(g,\tau)$ is known for now and obtain the necessary conditions for stationarity:
    \begin{alignat}{2}\label{eq:snapshots_variation_UL_eta}
       \partial L^\text{UL}/\partial {}^\tau\!\eta_c^\alpha = 0:& \qquad {}^\tau\!\!f_c^\alpha - \sum_g w^g {}^\tau\!\!J^g_{\!\!\vartriangle} {}^\tau\!h^g {}^\tau\!b_{\!abc}^{\alpha g} {}^\tau\!\sigma_{ab}^g = 0 \; ,
       \\ \label{eq:snapshots_variation_UL_mu}
       \partial L^\text{UL}/\partial  {}^\tau\!\mu_c^\beta  = 0:& \qquad {}^\tau\!\zeta_c^\beta - \sum_g w^g {}^\tau\!\!J^g_{\!\!\vartriangle} {}^\tau\!h^g {}^\tau\!b_{\!abc}^{\beta g} {}^\tau\!\sigma_{ab}^g = 0 \; ,
       \\ \label{eq:snapshots_variation_UL_zeta}
       \partial L^\text{UL}/\partial  {}^\tau\!\zeta_c^\beta  = 0:& \qquad  {}^\tau\!\mu_c^\beta = 0 \; ,
       \\ \label{eq:snapshots_variation_UL_stress}
       \partial L^\text{UL}/\partial {}^\tau\!\sigma_{ab}^g  = 0:& \qquad  
       \begin{aligned}[t]
       & w^g {}^\tau\!\!J^g_{\!\!\vartriangle} {}^\tau\!h^g c_{abcd}^{-1} \left({}^\tau\!\sigma_{cd}^g - \sigma_{cd}^{*s(g,\tau)}\right)\\
       - \;\; & w^g {}^\tau\!\!J^g_{\!\!\vartriangle} {}^\tau\!h^g \bigg(\!\!\!\!\sum_{\substack{\alpha \\ \pi(\alpha) = 1}}\!\!\!\! {}^\tau\!b_{\!abc}^{\alpha g} {}^\tau\!\eta_c^\alpha + \!\!\!\!\sum_{\substack{\beta \\ \pi(\beta) = 0}}\!\!\!\! {}^\tau\!b_{\!abc}^{\beta g} {}^\tau\!\mu_c^\beta\bigg) = 0\; ,
       \end{aligned}
       \\ \label{eq:snapshots_variation_UL_strain_mat}
       \partial L^\text{UL}/\partial e_{kl}^{*z}  = 0:& \qquad -\sum_\tau \sum_g w^g {}^\tau\!\!J^g_{\!\!\vartriangle} {}^\tau\!h^g c_{klmn} \left({}^\tau\!e_{mn}^g - e_{mn}^{*s(g,\tau)}\right) \delta_{s(g,\tau)z} = 0 \; ,
       \\ \label{eq:snapshots_variation_UL_stress_mat}
       \partial L^\text{UL}/\partial \sigma_{ab}^{*z}  = 0:& \qquad -\sum_\tau \sum_g w^g {}^\tau\!\!J^g_{\!\!\vartriangle} {}^\tau\!h^g c_{abcd}^{-1} \left({}^\tau\!\sigma_{cd}^g - \sigma_{cd}^{*s(g,\tau)}\right) \delta_{s(g,\tau)z} = 0 \; .
    \end{alignat}
    
    We directly find the condition ${}^\tau\!\mu_c^\beta = 0$ for all global node numbers $\beta$ with $\pi(\beta) = 0$. By using this, we can rearrange the condition $\partial L^\text{UL}/\partial {}^\tau\!\sigma_{ab}^g  = 0$ to
    \begin{align}\label{eq:snapshots_equation_UL_update_stress}
    \eqref{eq:snapshots_variation_UL_stress}:& \qquad {}^\tau\!\sigma_{ab}^g = \sigma_{ab}^{*s(g,\tau)} + \!\!\!\!\sum_{\substack{\alpha \\ \pi(\alpha) = 1}}\!\!\!\!c_{abcd} {}^\tau\!b_{\!cde}^{\alpha g} {}^\tau\!\eta_e^\alpha \; .
    \end{align}
    By rearranging $\partial L^\text{UL}/\partial e_{kl}^{*z}  = 0$ we get
    \begin{align}
    \label{eq:snapshots_equation_UL_update_strain}
       \eqref{eq:snapshots_variation_UL_strain_mat}:& \qquad
       e_{kl}^{*z} = \dfrac{1}{\sum\limits_\tau\sum\limits_g w^g {}^\tau\!\!J^g_{\!\!\vartriangle} {}^\tau\!h^g \delta_{s(g,\tau)z}} \sum_\tau\sum_g w^g {}^\tau\!\!J^g_{\!\!\vartriangle} {}^\tau\!h^g {}^\tau\!e_{kl}^g \delta_{s(g,\tau)z} \; .
    \end{align}
    By inserting into each other and rearranging, we also obtain
    \begin{align}
       \label{eq:snapshots_equation_UL_system_1}
       \eqref{eq:snapshots_variation_UL_stress} \; \text{in} \; \eqref{eq:snapshots_variation_UL_stress_mat}:& \qquad - \sum_\tau\!\!\!\sum_{\substack{\alpha \\ \pi(\alpha) = 1}}\!\!\!\sum_g w^g {}^\tau\!\!J^g_{\!\!\vartriangle} {}^\tau\!h^g {}^\tau\!b_{abc}^{\alpha g} {}^\tau\!\eta_c^\alpha \delta_{s(g,\tau)z} = 0 \text{ and }
       \\ \label{eq:snapshots_equation_UL_system_2}
       \eqref{eq:snapshots_equation_UL_update_stress} \; \text{in} \; \eqref{eq:snapshots_variation_UL_eta}:& \qquad \!\!\!\!\sum_{\substack{\beta\\ \pi(\beta) = 1}}\!\!\!\! \sum_g w^g {}^\tau\!\!J^g_{\!\!\vartriangle} {}^\tau\!h^g {}^\tau\!b_{abc}^{\alpha g} c_{abde} {}^\tau\!b_{de\!f}^{\beta g} {}^\tau\!\eta_{\!f}^\beta + \sum_g w^g {}^\tau\!\!J^g_{\!\!\vartriangle} {}^\tau\!h^g {}^\tau\!b_{abc}^{\alpha g} \sigma_{ab}^{*s(g,\tau)} = {}^\tau\!\!f_c^\alpha \; .
    \end{align}
    To solve the set of equations \eqref{eq:snapshots_variation_UL_mu} and \eqref{eq:snapshots_equation_UL_update_stress}--\eqref{eq:snapshots_equation_UL_system_2} for ${}^\tau\!\zeta_a^\beta$, ${}^\tau\!\sigma_{ab}^g$, $e_{kl}^{*z}$, $\sigma_{ab}^{*z}$ and ${}^\tau\!\eta_a^\alpha$ and to determine the discrete mapping $s(g,\tau)$, we apply a decoupled algorithm in an adapted version of \cite{Leygue2018}, cf.~Alg.~\ref{alg:data-driven_updated_Lagrange}. For further details on the decoupled algorithm, the reader is referred to Remark~\ref{remark:algorithm}.
    In the case of anisotropic materials, the updated Lagrangian DDI formulation based on $(\te e, \teg \sigma)$ requires interpolation-based postprocessing, which then allows the generated material database to be used for training a PANN formulated with structure tensors and invariants, cf. Remark~\ref{remark:anisotropic}.

    \subsubsection{Total Lagrangian formulations}\label{sec:ddi_total_lagrange}

    As an alternative to the updated Lagrangian DDI formulation, we introduce two \emph{total Lagrangian DDI formulations}, one called \emph{original} and one called \emph{adapted}. 
    To start with, we could use either the conjugate pair $(\te F, \te P)$ or $(\te E, \te T)$. Since a formulation with the 2nd Piola-Kirchhoff tensor allows the angular momentum balance to be fulfilled by construction, the second formulation is to be preferred.
    
    Compared to the updated Lagrangian DDI formulation based on $(\te e,\teg \sigma)$, this formulation based on $(\te E,\te T)$  offers the advantage that the material database to be generated can be directly used for training a PANN for anisotropic materials formulated using structure tensors and invariants, cf. Remark~\ref{remark:anisotropic}.
    
    \paragraph{Original total Lagrangian DDI formulation}
    The discrete loss term for the original total Lagrangian DDI formulation can be derived in a similar way as described in \ref{app:discrete_ddi}. In contrast to the updated Lagrangian DDI formulation, the integration domain is here $\mathcal B_0$. Thus, the loss term follows to
    \begin{align}\label{eq:snapshots_loss_total_lagrange_main}
    \begin{split}
        L^\text{TL} := &\frac{1}{2} \sum_\tau\sum_g w^g J^g_{0,\vartriangle} h^g_{0,\vartriangle} \biggl[\left({}^\tau\!E_{\!K\!L}^g - E_{\!K\!L}^{*s(g,\tau)}\right) C_{\!K\!L M\!N} \left({}^\tau\!E_{\!M\!N}^g - E_{\!M\!N}^{*s(g,\tau)}\right) 
        \\
        &+ \left({}^\tau T_{\!AB}^g - T_{\!AB}^{*s(g,\tau)}\right) C_{\!ABC\!D}^{-1} \left({}^\tau T_{\!C\!D}^g - T_{\!C\!D}^{*s(g,\tau)}\right)\biggr] 
        \\
        &+  \sum_\tau\!\!\!\!\sum_{\substack{\alpha \\ \pi(\alpha) = 1}}\!\!\!\! {}^\tau\!\eta_c^\alpha \bigg({}^\tau\!\!f_c^\alpha - \sum_g w^g J^g_{0,\vartriangle} h^g_{0,\vartriangle} {}^\tau\!B_{\!ABc}^{\alpha g} {}^\tau T_{\!AB}^g\bigg)
        \\
        &+  \sum_\tau\!\!\!\!\sum_{\substack{\beta\\ \pi(\beta) = 0}}\!\!\!\! {}^\tau\!\mu_c^\beta \bigg({}^\tau\!\zeta_c^\beta - \sum_g w^g J^g_{0,\vartriangle} h^g_{0,\vartriangle} {}^\tau\!B_{\!ABc}^{\beta g} {}^\tau T_{\!AB}^g\bigg)
    \end{split} \; ,
    \end{align}
    where the pseudo stiffness tensor is given by $\tte C = C \ttes 1$ with $C \in \R_{>0}$. Since we integrate over the elements with reference domain $\Omega_0^e$, cf. Eq.~\eqref{eq:finite_elements}, the determinant of the Jacobian matrix $J^g_{0,\vartriangle}$ in the undeformed state is used. The solution strategy is completely analogous to the updated Lagrangian DDI formulation.

    \begin{figure}[t!]
    	    \centering
    	    \includegraphics[width=\textwidth]{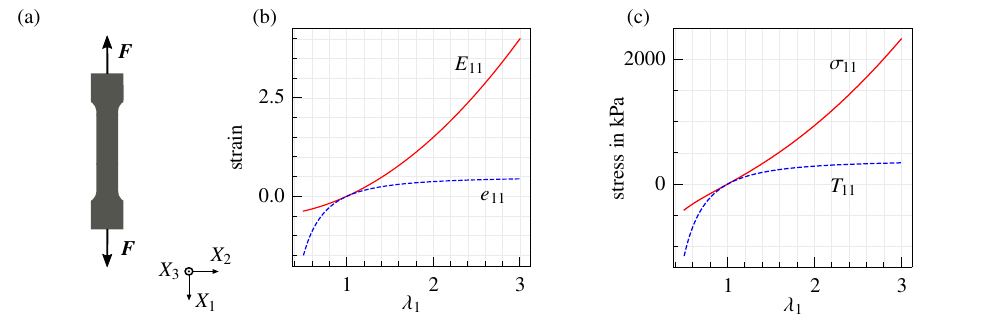}
    	    \caption{Uniaxial tensile test of compressible neo-Hooke material with $E:=\SI{1}{\mega\pascal}, \nu := 0.3$: (a) homogenous specimen under loading conditions and $11$-component of (b) Euler-Almansi and Green-Lagrange as well as (c) 2nd Piola-Kirchhoff and Cauchy stress tensor.
    	    }
    	    \label{fig:uniaxial_tensile_test}
    	\end{figure}

    \paragraph{Adapted total Lagrangian DDI formulation}
    In addition to the formulation given by the loss term~\eqref{eq:snapshots_loss_total_lagrange_main}, we introduce an \emph{adapted total Lagrangian DDI formulation} in the following. The idea here is to no longer assume that the pseudo stiffness is constant in space and time. 
    Instead, it is defined via the pull-back of the pseudo stiffness tensor $\ttes c = C \ttes 1$, which corresponds to $\teg \sigma$ in the incremental relation $\overset{\circ}{\teg \sigma} = \ttes c : \te d$ with $\overset{\circ}{\teg \sigma}$ the Truesdell rate and $\te d = \sym(\dot{\te F} \cdot \te F^{-1})$, i.e., 
    \begin{align}
    \label{eq:pull_back_stiffness}
        \tte C^\text{ada} := \mathrm{pb}(C \ttes 1, \te F) &= \frac{C}{2} (\delta_{km}\delta_{ln} + \delta_{kn}\delta_{lm}) JF_{Kk}^{-1}F_{Ll}^{-1}F_{Mm}^{-1}F_{Nn}^{-1}  \ve e_K \otimes \ve e_L \otimes \ve e_M \otimes \ve e_N \\
        &= \frac{C J}{2} \left(C_{KM}^{-1} C_{LN}^{-1} + C_{KN}^{-1} C_{LM}^{-1}\right) \ve e_K \otimes \ve e_L \otimes \ve e_M \otimes \ve e_N \label{eq:pull-back-c}
    \end{align}
    with $C \in \R_{>0}$. 
    This idea stems from the fact that the stress-strain curves for the 2nd Piola-Kirchhoff stress tensor $\te T$ are significantly more nonlinear as a result of the pull-back operation from $\teg \sigma$ to $\te T$ and therefore the true or ground truth stiffness $\tte C^\text{gt}(\te F)$ of the material also changes significantly as a result of deformation, cf.~Fig.~\ref{fig:uniaxial_tensile_test}. This fact is taken into account by using the adapted pseudo-stiffness according to Eq.~\eqref{eq:pull-back-c}.
    The only change in Eq.~\eqref{eq:snapshots_loss_total_lagrange_main} that needs to be made is therefore to replace $C_{K\!L M\!N}$ with ${}^\tau C_{K\!L M\!N}^{\text{ada},g}$ which is a function of the deformation gradient ${}^\tau \!F_{kK}^g$ at snapshot $\tau$ and the quadrature point $g$. 
    
    Due to the fact that the pseudo stiffness is not assumed to be a constant anymore, the step for calculating the material states $\te E^{*z}$ has to be modified according to
    \begin{align}
        E_{K\!L}^{*z} = \bigg(\sum\limits_\tau\sum\limits_g w^g J^g_{0,\vartriangle} h^g_{0,\vartriangle} {}^\tau C_{K\!L M\!N}^g\delta_{s(g,\tau)z}\bigg)^{-1} \sum_\tau\sum_g w^g J^g_{0,\vartriangle} h^g_{0,\vartriangle} {}^\tau C_{\!M\!N\!PQ}^g {}^\tau\!E_{PQ}^g \delta_{s(g,\tau)z} \; .
    \end{align}
    The remaining solution algorithm stays untouched, cf.~Alg.~\ref{alg:data-driven_adapted_total_Lagrange}.

    \subsection{Physics-augmented neural networks}\label{sec:neural_networks}
    In this section, we present a \emph{polyconvex PANN} for modeling the constitutive behavior of \emph{isotropic elastic} solids, as originally developed in Linden~et~al.~\cite{Linden2023}.
    
    \subsubsection{Model formulation for isotropic elasticity}\label{subsec:PANN_model}
    \begin{figure}[b!]
    	    \centering
    	    \includegraphics[width=1.\textwidth]{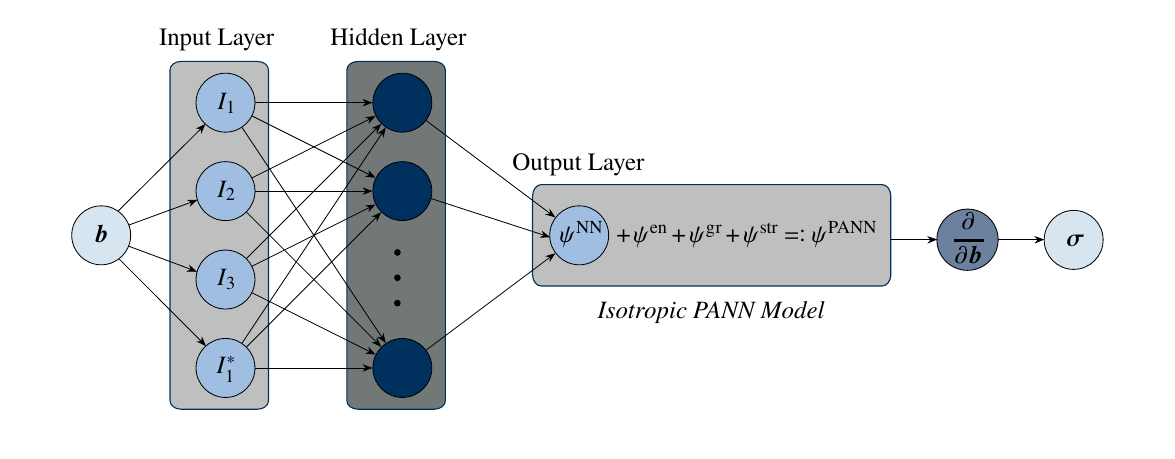}
    	    \caption{Illustration of the PANN for modeling the constitutive behavior of isotropic elastic solids. Note that the hidden layer of the
ICNN may be multilayered.
    	    }
    	    \label{fig:neural_network}
    	\end{figure}
    The isotropic hyperelastic material behavior can be described by the complete and irreducible set of invariants $\boldsymbol{\mathcal I} := (I_1, I_2, I_3)$, cf. Sect.~\ref{sec:basics}.
    However, it may be convenient to incorporate the additional invariant $I_1^* := -2J$ into the argument list of the predicted potential, e.g., to improve the approximation quality of the predictions~\cite{Klein2021,Linden2023}.
    By utilizing the extended set of invariants $\boldsymbol{\mathcal I}^* := (I_1, I_2, I_3, I_1^*)$ as inputs to an FICNN with only one hidden layer including $N^\text{NN}$ neurons and a scalar-valued output
    \begin{align}
    \label{eq:NN_model}
    \psi^{\text{NN}}(\boldsymbol{\mathcal I}^*) :=  
        \sum_{\alpha=1}^{N^\text{NN}} W_{\alpha}\,\Softplus\left(\sum_{\beta=1}^{3} w_{\alpha\beta} I_\beta + w_{\alpha 1}^* I_1^* + b_\alpha\right)\, ,
    \end{align}
    which is then interpreted as a \emph{hyperelastic potential}, the flexibility of the resulting model significantly surpasses that of classical model formulations.
    Here, the \emph{Softplus} activation function $\Softplus(x):=\log(1+\exp(x))\in\mathcal{C}^{\infty}$ is applied, which is convex and non-decreasing.
    The weights $W_{\alpha}, w_{\alpha\beta} \in \R_{\ge 0},  w_{\alpha 1}^* \in \R$ and bias values $b_\alpha\in\R$ together form the set of parameters $\boldsymbol{\mathcal P} \in \R^P$, with $P\in \N$ denoting the total number of parameters, to be optimized in the calibration process to fit a given dataset, cf. Section~\ref{subsec:PANN_training}. Due to the chosen activation function and the weight restrictions, the model is \emph{polyconvex} by construction \cite{Klein2021,Linden2023}.
    
    Following \cite{Linden2023}, the isotropic PANN model is introduced by the potential
    \begin{align}
    \label{eq:PANN_model}
    \psi^\text{PANN}(\boldsymbol{\mathcal I}^*) := \psi^\text{NN} + \psi^\text{en} + \psi^\text{gr}(J) + \psi^\text{str}(J)
    \end{align}
    with the correction terms
    \begin{align}
    \label{eq:PANN_analytical_terms}
    \psi^\text{en} := -\psi^\text{NN}(\boldsymbol{\mathcal I}^*)\big\vert_{\te F =  \te 1}\,,
    \quad
    \psi^\text{gr}(J) := \lambda_\text{gr}\Big(J + \frac{1}{J} - 2\Big)^2\,, 
    \quad
    \psi^\text{str}(J) := - \mathfrak{n}(J - 1)\, ,
    \end{align}
    guaranteeing \emph{energy- and stress normalization} as well as \emph{volumetric growth condition}, 
    and the isotropic stress normalization constant
    \begin{align}
    \label{eq:PANN}
    \mathfrak{n} = 2 \,\bigg(\,\frac{\partial \psi^{\text{NN}}}{\partial I_1}
     +
     2\frac{\partial \psi^{\text{NN}}}{\partial I_2}
    +
     \frac{\partial \psi^{\text{NN}}}{\partial I_3}
    +\frac{\partial \psi^{\text{NN}}}{\partial I_1^*} \frac{\partial I_1^*}{\partial I_3}
     \bigg)\Bigg\rvert_{\te F = \te 1} \in \R \; .
    \end{align}
    In Eq.~\eqref{eq:PANN_analytical_terms}, the parameter $\lambda_\text{gr}\in\R_{>0}$ has to be chosen such that the energy grows fast enough during compression. Following \cite{Kalina2024}, a value between \num{1e-2} and \num{1e-3} the material’s initial stiffness has proven to be suitable.
    Then, the model defined in Eq.~\eqref{eq:PANN_model} fulfills all mentioned requirements from Sect.~\ref{sec:Hyper} by construction except for the non-negativity of the energy function. A schematic representation of the PANN model is given in Fig.~\ref{fig:neural_network}.
    The stress tensors $\te T$ or $\teg \sigma$ following from the overall potential $\psi^\text{PANN}(\boldsymbol{\mathcal I}^*)$ can be obtained using Eq.~\eqref{eq:stress_invariants}.

    \subsubsection{Calibration of the PANN model}\label{subsec:PANN_training}
  
    In order to calibrate the proposed model for a specific material, datasets of strain-stress pairs either in terms of the right Cauchy-Green deformation tensor $\te C$ and the second Piola-Kirchhoff stress tensor $\te T$ or the left Cauchy-Green deformation tensor $\te b$ and the Cauchy stress tensor $\teg \sigma$ are used. 
    Then, the dataset $\mathcal{D}$ is split into a calibration set $\mathcal D_\text{c}$ and a test set $\mathcal{D}_\text{t}$, such that $\mathcal D = \mathcal D_\text{c} \cup \mathcal D_\text{t}$ and $\mathcal{D}_\text{c} \cap \mathcal{D}_\text{t} = \varnothing$ hold. 
    Finally, model generalization is evaluated on $\mathcal{D}_\text{t}$ after calibration on $\mathcal{D}_\text{c}$. Only if predictions for unseen data are accurate, the model can be assumed to generalize well.
    
    The model architecture, including hyperparameters like the number of layers and nodes, must be chosen sufficiently large to capture the material behavior. Overfitting, a common challenge for neural networks, is mitigated in the proposed PANN model due to its inherent mathematical structure, cf. Section~\ref{subsec:PANN_model}.
    Model parameters $\boldsymbol{\mathcal P}$ consisting of weights and biases are optimized via the SLSQP optimizer by minimizing the loss function defined as the mean squared error:
    \begin{align}
    \label{eq:mse_calibration}
     \mathscr{M\!S\!E}^\text{UL}(\boldsymbol{\mathcal P})=\frac{1}{\vert\mathcal{D}_\text{c}\vert}\sum_{i = 1}^{\vert\mathcal{D}_\text{c}\vert} \Big\Vert{}^i\teg \sigma-\teg \sigma^{\text{PANN}}({}^i\te b;\,\boldsymbol{\mathcal P})\Big\Vert^2 \quad \text{or} \quad \mathscr{M\!S\!E}^\text{TL}(\boldsymbol{\mathcal P})=\frac{1}{\vert\mathcal{D}_\text{c}\vert}\sum_{i = 1}^{\vert\mathcal{D}_\text{c}\vert} \Big\Vert{}^i\te T-\te T^{\text{PANN}}({}^i\te C;\,\boldsymbol{\mathcal P})\Big\Vert^2\; ,
    \end{align}
    where $\Vert\cdot\Vert$ denotes the Frobenius norm. \emph{Sobolev training} is applied, meaning the hyperelastic potential $\psi^\text{PANN}$ is only calibrated based on the derivative, i.e., stress values, rather than functional values. The implementation leverages \emph{Python, TensorFlow}, and \emph{SciPy}.
    
    
    \section{Numerical examples}\label{sec:examples}

    Within this section, we demonstrate the ability of the developed \emph{data-driven dual-stage approach} for the automated generation of constitutive models. In the absence of real experiments, the input data for the DDI is initially generated via \emph{virtual experiments} using FE simulations within this work. We look at two different examples: One in which all required quantities for the DDI, i.e., displacement field and external nodal forces, are known and noise-free and one in which the situation of a real experiment is mimicked. In the first example, only the DDI step is considered and the three formulations are benchmarked, while in the second example the entire dual-stage framework is tested.

    \subsection{Ground truth constitutive model}

    To mimic a real experiment with an FE simulation, a constitutive model has to be chosen. This model serves as the ground truth for the DDI as well as for the trained PANN including 3D FE simulations.
    Herein, to describe the behavior of a compressible elastic material, we choose a two-parametric neo-Hookean model
    \begin{align}
    \label{eq:energy_NH}
        \psi^{\text{nh}}(I_1,I_3)=\frac{1}{2}\left(\mu\left(I_1-\ln I_3 - 3\right)+\frac{\lambda}{2}\left(I_3-\ln I_3 -1\right)\right)\,, \quad \mu = \frac{E}{2(1 + \nu)}\,, \quad \lambda = \frac{E \nu}{(1 + \nu)(1 - 2\nu)}
    \end{align}
    according to Ciarlet~\cite{Ciarlet1988}, with the material parameters $(E,\nu)$ corresponding to the initial Young's modulus and Poisson's ratio.
    This model fulfills all mentioned requirements from Sect.~\ref{sec:Hyper} and from this potential, the 2nd Piola-Kirchhoff stress and the Cauchy stress can be derived via Eq.~\eqref{eq:stress_invariants} as 
    \begin{align}
    \label{eq:stress_NH}
        \te T^{\text{nh}} = \mu\te 1 + \left(\frac{\lambda}{2}-\frac{2\mu+\lambda}{2I_3}\right)\cof \te C 
        \quad \text{and} \quad
        \teg \sigma^{\text{nh}} = \frac{\mu}{J}\te b + \left(\frac{\lambda J}{2}-\frac{2\mu+\lambda}{2J}\right) \te 1 \; , 
    \end{align}
    respectively. The material parameters are chosen to $E:=\SI{1}{\mega\pascal}$ and $\nu := 0.3$ within all numerical examples presented in the following. Stress-stretch curves are shown for a uniaxial compression/tension test within Fig.~\ref{fig:uniaxial_tensile_test}, where also the strain measures $\te E$ and $\te e$ are depicted.

    \subsection{Benchmarking the DDI formulations with ideal data}\label{subsec:benchmarks}
    In this first example, we compare the three developed DDI formulations, i.e., the \emph{updated Lagrangian formulation}, the \emph{total Lagrangian formulation} and the \emph{adapted total Lagrangian formulation}, in a \emph{benchmark test} with respect to their accuracy. To evaluate the accuracy for both, the mechanical states $\left({}^\tau \!\te e^g,{}^\tau \!\teg\sigma^g\right)$ or $\left({}^\tau \te E^g,{}^\tau \te T^g\right)$ and the material states $\left(\te e^{*z},\teg \sigma^{*z}\right)$ or $\left(\te E^{*z},\te T^{*z}\right)$, the neo-Hookean ground truth model \eqref{eq:energy_NH} is used. We consider the strain in the tuples to be fixed and calculate the associated stress with Eq.~\eqref{eq:stress_NH}${}$, i.e.,
    \begin{align}
    {}^\tau\!\teg \sigma^{g,\text{nh}}=\teg \sigma^\text{nh}({}^\tau\te b^g) 
    \quad \text{and} \quad 
    {}^\tau\te T^{g,\text{nh}}=\te T^\text{nh}({}^\tau\te C^g) \; ,
    \label{eq:ref_stress}
    \end{align}
    where the left Cauchy-Green deformation and the
    right Cauchy-Green deformation are calculated by ${}^\tau\te b^g = \left(\te 1 - 2 {}^\tau\!\te e^g\right)^{-1}$ and ${}^\tau\te C^g = 2 {}^\tau\te E^g + \te 1$, respectively. The same procedure is applied to the material states, e.g., $\teg \sigma^{*z,\text{nh}}=\teg \sigma^\text{nh}(\te b^{*z})$ as well as $\te T^{*z,\text{nh}}=\te T^\text{nh}(\te E^{*z})$.

    The generated data of this benchmark will be made freely available in the final version of the article.
    
    \subsubsection{Setup}

    \begin{figure}
    	    \centering
    	    \includegraphics[width=\textwidth]{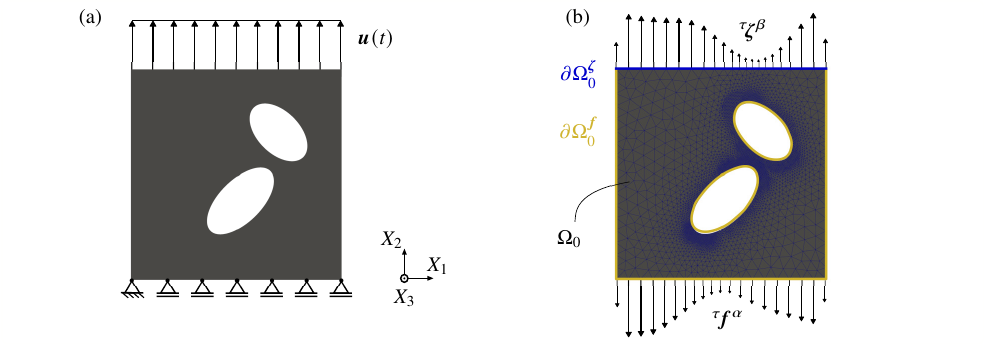}
    	    \caption{Benchmark test: (a) boundary conditions with $u_2^\text{max} = \SI{50}{\milli\meter}$ for uniaxial tensile test of specimen with reference thickness $h_0 = \SI{5}{\milli\meter}$ to be investigated as well as (b) domain of interest, given by $\SI{100}{\milli\meter} \times \SI{100}{\milli\meter} \times \SI{5}{\milli\meter}$, for DDI, which contains the interior $\Omega_0$ with the two ellipsoidal holes as well as the boundaries of prescribed and unknown nodal forces.
    	    }
    	    \label{fig:benchmark}
    \end{figure}

    We consider a thin plate with two ellipsoidal holes, loaded with displacement boundary conditions (BCs), see Fig.~\ref{fig:benchmark}. The mesh consist of $N^\text{nodes}=4403$ nodes and $N^\text{quad}=8444$ quadrature points.
    Within the benchmark test, we use ideal, non-noisy field quantities from a \emph{2D plane stress FE simulation} with $N^\text{snap} = 10$ increments and provide all required input data for the DDI, i.e., the exact in-plane nodal displacements ${}^\tau\! u_1^\alpha$, ${}^\tau\! u_2^\alpha$ and the external forces ${}^\tau\!\!\ve f^\alpha$ at the nodes $\alpha \in \{1,2,\dots N^\text{node}\}$ of the mesh are available for each snapshot $\tau\in\{1,\dots,N^\text{snap}\}$. We also take the thickness ${}^\tau\! h^g \in \R_{>0}$ at the quadrature points $g\in\{1,\dots,N^\text{quad}\}$ of the deformed elements and the reference thickness $h_0 = \SI{1}{\milli\meter}$ as given. Note that this quantity is needed if a compressible material is investigated, cf. Remark~\ref{remark:compressible}.  
    Thus, for now, we will ignore the fact that in real experiments only the global testing force would be known and the corresponding external nodal forces can only be calculated under certain conditions.
    The lower boundary of the domain $\mathcal B_0$ is part of $\partial \mathcal B_0^{\ve f}$, i.e., the external nodal forces ${}^\tau\!\!\ve f^\alpha$ are given, and the upper boundary is set to $\partial \mathcal B_0^{\ve \zeta}$, i.e., the external nodal forces labeled ${}^\tau\!\ve \zeta^\alpha$ are not given and can be calculated as part of the DDI, if requested.

    \subsubsection{Results}
    \paragraph{Fixed hyperparameters}

    \begin{figure}
    	    \centering
    	    \includegraphics[width=\textwidth]{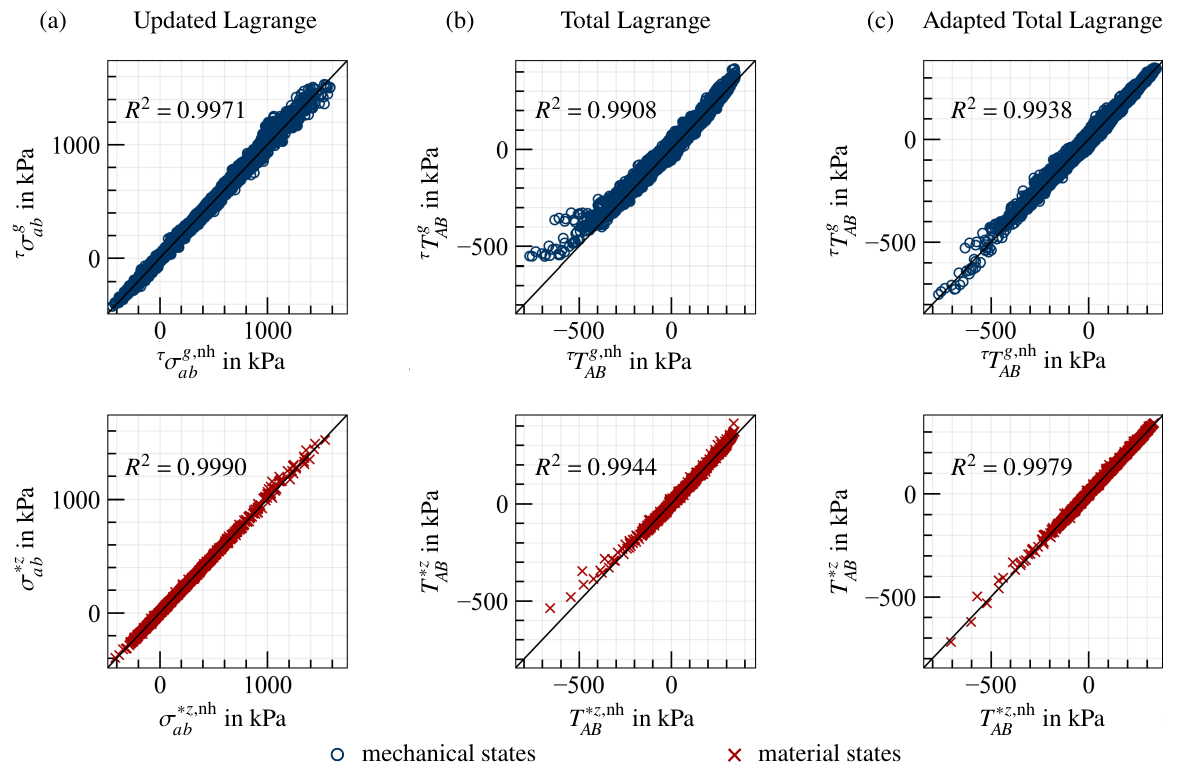}
    	    \caption{Quantitative comparison of Neo-Hooke reference model and DDI results obtained with (a) Updated-Lagrange-, (b) Total-Lagrange- and (c) Adapted-Total-Lagrange formulation: mechanical states and material states based on ideal data from perforated disc discretized with $N^\text{quad} = 8444$ linear elements under loading in each snapshot $\tau \in \{1, \dots, N^\text{snap}\}$ with $N^\text{snap} = 10$ as well as $N^* := \lceil  0.01 \cdot N^\text{quad}\cdot N^\text{snap}\rceil = 845$ as prescribed number of material states of and the given pseudo stiffness tensor $\mathbb{C} := \SI{1}{\mega\pascal} \,\mathbbm 1$.
    	    }
    	    \label{fig:ideal_data_plot_states}
    	\end{figure}
    \begin{figure}
    	    \centering
    	    \includegraphics[width=\textwidth]{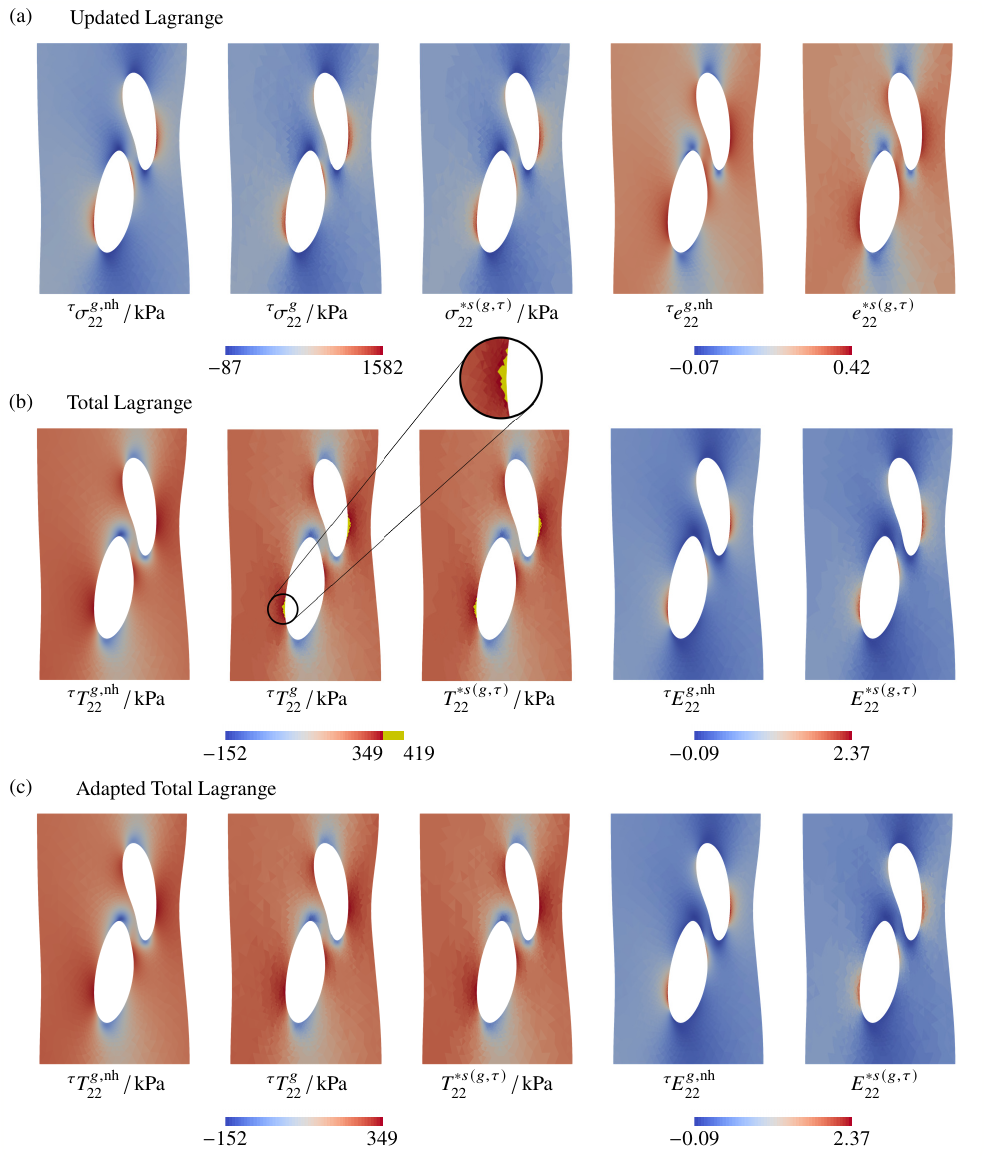}
    	    \caption{Qualitative comparison of Neo-Hooke reference model and DDI field data obtained with (a) Updated-Lagrange-, (b) Total-Lagrange- and (c) Adapted-Total-Lagrange formulation: mechanical states and material states based on ideal data from perforated disc discretized with $N^\text{quad} = 8444$ linear elements under loading in the last snapshot $\tau = 10$ as well as $N^* := \lceil  0.01 \cdot N^\text{quad}\cdot N^\text{snap}\rceil = 845$ as prescribed number of material states of and the given pseudo stiffness tensor $\mathbb{C} := \SI{1}{\mega\pascal} \,\mathbbm 1$.
    	    }
    	    \label{fig:ideal_data_plot_paraview}
    	\end{figure}

    First, we start with the following hyperparameters: The number of material states is chosen to $N^* := \lceil  0.01 \cdot N^\text{quad}\cdot N^\text{snap}\rceil = 845$ and the pseudo stiffness tensors $\ttes c := \SI{1}{\mega\pascal} \,\mathbbm 1$ or $\tte C := \SI{1}{\mega\pascal} \,\mathbbm 1$. 

    A comparison of the mechanical and material states generated with the \emph{updated Lagrangian DDI formulation} is provided in Fig.~\ref{fig:ideal_data_plot_states}(a), where the stress ${}^\tau\!\teg \sigma^{g,\text{nh}}$ as defined in Eq.~\eqref{eq:ref_stress}${}_1$ is used for evaluation. Acceptable accuracy can already be seen in the mechanical states. However, these states are not used as a material database for training the PANN, but only the material states, cf. Sect.~\ref{sec:dual_stage}. For the material states, the deviation from the ground truth model is then only slight and a very good $R^2$ value of $0.999$ is achieved. In Fig.~\ref{fig:ideal_data_plot_paraview}(a), surface plots of the true stress field ${}^{10}\sigma_{22}^{g,\text{nh}}$, the mechanical states ${}^{10}\sigma_{22}^{g,\text{nh}}$ and the assigned material states $\sigma_{22}^{*s(g,10)}$ as well as the true strain field ${}^{10} e_{22}^{g,\text{nh}}$ and the assigned material states $e_{22}^{*s(g,10)}$ are shown for the last snapshot $\tau=10$. Here, too, it can be seen that a very high level of accuracy can be achieved.

    In Fig.~\ref{fig:ideal_data_plot_states}(b), the results for the \emph{total Lagrangian DDI formulation} are shown, where the stress ${}^\tau\te T^{g,\text{nh}}$ as defined in Eq.~\eqref{eq:ref_stress}${}_2$ is used here. Compared to the updated Lagrangian formulation, there is a significant decrease in accuracy. This is particularly noticeable for large compressive and tensile stresses. We assume that this is due to the fact that the pull-back operation $\te T = J \te F^{-1} \cdot \teg \sigma \cdot \te F^{-T}$ results in a much more pronounced non-linear stress-deformation behavior, see Fig.~\ref{fig:uniaxial_tensile_test}(c). As shown in \cite{Kalina2022a}, a similar behavior can also be observed for a more complex Ogden model. This can also be seen in the surface plots given in Fig.~\ref{fig:ideal_data_plot_paraview}(b). These are scaled according to the reference solution and areas with values outside the scale are colored yellow. It can be seen that areas with overestimated stresses occur at the edges of the holes for the mechanical states ${}^{10}\sigma_{22}^{g,\text{nh}}$ and the assigned material states $\sigma_{22}^{*s(g,10)}$.

    Finally, Fig.~\ref{fig:ideal_data_plot_states}(c) shows the results obtained with the \emph{adapted total Lagrangian DDI formulation}. As described in Sect.~\ref{sec:ddi_total_lagrange}, the idea of this formulation is to set the pseudo stiffness $\ttes c =\text{const.}$ and to use an adapted pseudo stiffness $\tte C^\text{ada}$ defined via the pull-back transformation of $\ttes c$ to $\tte C^\text{ada}(\ttes c, \te F)$ given in Eq.~\eqref{eq:pull_back_stiffness}, i.e., the pseudo stiffness $\tte C^\text{ada}$ related to $\te T$ is therefore not assumed to be constant but a function of the deformation gradient ${}^\tau \te F^g$ varying at each quadrature point $g$ and each snapshot $\tau$. Thus, the pseudo stiffness is adapted to the strongly non-linear course of the stress-stretch relationship for $\te T$ depending on the deformation state, which results in a significant improvement of the generated database. In particular, the deviations of the original total Lagrangian DDI formulation at high compressive and tensile stresses are noticeably corrected. As a result, the yellow-colored areas with overestimated stresses ${}^{10}\sigma_{22}^{g,\text{nh}}$ and $\sigma_{22}^{*s(g,10)}$  no longer appear in the surface plots in Fig.~\ref{fig:ideal_data_plot_paraview}(c).

    \paragraph{Hyperparameter study}

    After considering fixed hyperparameters, we carry out a parameter variation to investigate the influence on the three finite strain DDI formulations. 
    
\begin{figure}
    	    \centering
    	    \includegraphics[width=\textwidth]{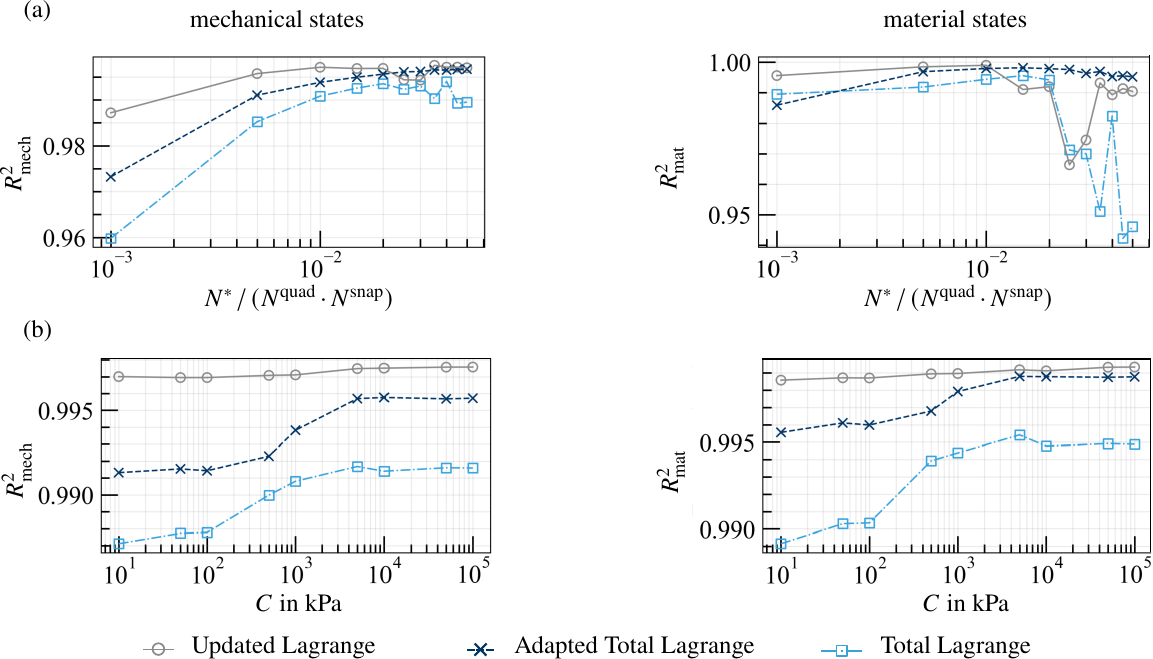}
    	    \caption{Parameter study of DDI with coefficient of determination $R^2$ for mechanical and material states depending on (a) the ratio of $N^*$ and number of quadrature points $N^\text{quad}$ times the number of snapshots $N^\text{snap}$ with prescribed pseudo stiffness tensor $\mathbb{C} := \SI{1}{\mega\pascal} \,\mathbbm 1$ as well as (b) the pseudo stiffness $C$ with number $N^* := \lceil  0.01 \cdot N^\text{quad}\cdot N^\text{snap}\rceil$ of material states.
    	    }
    	    \label{fig:ideal_data_plot_r2_stiff}
    	\end{figure}

    First, we fix the pseudo stiffness to $C=10^3\,\text{kPa}$ and vary the ratio $N^*/(N^\text{quad}\cdot N^\text{snap})\in \{\SI{1e-3}{},\SI{5e-3}{}, \SI{1e-2}{}, \dots, \SI{5e-2}{}\}$, i.e., the number of material states with respect to number of quadrature points times snapshots.
    From Fig.~\ref{fig:ideal_data_plot_r2_stiff}(a), it can be seen that in the mechanical states, all three formulations result in an increase in accuracy for a larger ratio $N^*/(N^\text{quad}\cdot N^\text{snap})$. 
    This can be explained by the fact that if the number of $N^*$ is too small, an inhomogeneous stress field, which is approximated by the mechanical states ${}^\tau \!\teg \sigma^g$ or ${}^\tau \te T^g$, can no longer be represented with sufficient accuracy, as the material states are then very far away from the real stress field, within this example ${}^\tau\! \teg \sigma^{g,\text{nh}}$ or ${}^\tau \te T^{g,\text{nh}}$. Nevertheless, a very accurate material database can be generated even for a low ratio $N^*/(N^\text{quad}\cdot N^\text{snap})$, i.e., the $R^2$ value for the material states is close to $1$. However, it can be observed that a too high number of material states $N^*$ can lead to a drop in accuracy, at least for the updated Lagrangian and the total Lagrangian DDI formulation.
    This is due to the fact that it can lead to an insufficient regularization. This occurs because, with a higher number of material states, many states will be assigned to only a few quadrature points, which prevents the behavior from being adequately averaged. As a result, the mapping $s(g,\tau)$ is approaching injectivity, and the system fails to average over the different mechanical states sufficiently, introducing significant error. It is therefore crucial to select a value for $N^*$ that lies between these extremes. These findings are in agreement with those presented in~\cite{Dalemat2024, Leygue2018}.

    Secondly, we fix the number of material states to  $N^* := \lceil  0.01 \cdot N^\text{quad}\cdot N^\text{snap}\rceil$ and vary the pseudo stiffness tensors $\ttes c = C \mathbbm 1$ or $\tte C = C \mathbbm 1$ according to $C\in \left\{\SI{1e1}{}, \SI{5e1}{}, \SI{1e2}{}, \dots, \SI{1e5}{}\right\}\,\text{kPa}$. The resulting $R^2$ values depending on $C$ are shown in Fig.~\ref{fig:ideal_data_plot_r2_stiff}(b). Here, again, the stress tensors according to Eq.~\eqref{eq:ref_stress} are used as a reference. As can be seen, the influence of the pseudo stiffness on the accuracy is nearly negligible for the \emph{updated Lagrangian DDI formulation}. In contrast, a noticeable dependency on $\tte C$ or $\tte C^\text{ada}$ appears for both the \emph{total Lagrangian DDI formulation} and the \emph{adapted total Lagrangian DDI formulation}. However, from a pseudo-stiffness of $C= \SI{5e3}{\kilo\pascal}$, there is no longer any significant influence. As with the fixed hyperparameters, the best results can be achieved with the updated Lagrangian DDI formulation and there is a clear drop in accuracy for all calculations with the total Lagrangian DDI formulation. In contrast, with the adapted total Lagrangian DDI formulation, the loss of accuracy in the material states for a pseudo stiffness of $C= \SI{5e3}{\kilo\pascal}$ and above is very low.
    Furthermore, the pseudo stiffness plays a crucial role in the convergence of the method, i.e., increasing $C$ up to a certain limit leads to a reduction in error. By selecting a larger value of $C$, the correspondence between the material and mechanical states based on strain values is prioritized, which is advantageous since the strains are directly measured and, therefore, more reliable compared to stresses that evolve throughout the algorithm’s convergence. Finally, $N^*$ has a greater impact than $C$, i.e., an inappropriate choice of $N^*$ cannot be rectified by a favorable choice of $C$.
    These observations regarding the impact of the pseudo stiffness are consistent with those reported in~\cite{Dalemat2024}.

    \subsubsection{Assessment of the three DDI formulations and guide values for the hyperparameters}\label{sec:assessment_DDI}

    Basically, it can be stated that the highest accuracy in the generated data base can be achieved with the \emph{updated Lagrangian DDI formulation}. However, as stated in Remark~\ref{remark:anisotropic}, this formulation can only be used to generate a material data base for training a PANN if an isotropic material is considered. The \emph{original total Lagrangian DDI formulation} should not be used, as there are noticeable reductions in accuracy of the data. 
    If an anisotropic material is considered, the \emph{adapted total Lagrangian DDI formulation} should be used. This enables a similarly high accuracy as with the updated Lagrangian DDI formulation and at the same time the calibration of an anisotropic PANN model with the generated database.

    Based on our study, we can recommend a ratio of $N^*/(N^\text{quad}\cdot N^\text{snap}) = \SI{1}{\percent}$. Other authors give similar guideline values \cite{Leygue2018,Leygue2019,Zschocke2023,Dalemat2024}. For the pseudo-stiffness $\ttes c$ or $\tte C$, an increase in the estimated stiffness of the tested material by a factor of $10$ is recommended.


    \subsection{Application of the dual-stage framework in a realistic setup with noisy data}

    Within this numerical example, we mimic the situation of a realistic experiment for data mining, i.e., only the global force and the displacement field are accessible. Both quantities are subject to noise. These data are used as input for the dual-stage approach, i.e., DDI is applied and the generated material data base is used to train a PANN. Finally, the calibrated PANN model is applied in a 3D FE simulation.

    \subsubsection{Imitation of a realistic experiment for data mining}\label{subsec:num_exp}

    \begin{figure}
        \centering
        \includegraphics[width=\textwidth]{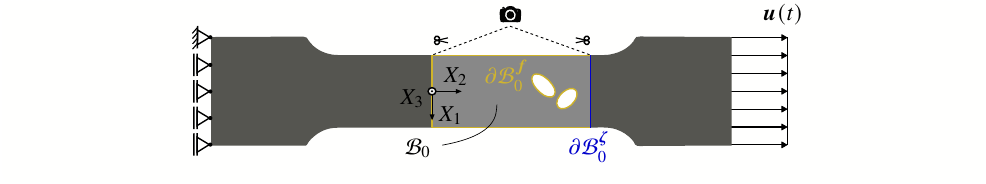}
        \caption{Realistic setup for uniaxial tensile test of specimen with reference thickness $h_0 = \SI{5}{\milli\meter}$ to be investigated in virtual experiment with clamping and displacement BCs with $u_2^\text{max} = \SI{210}{\milli\meter}$. The domain of interest, approximately $\SI{100}{\milli\meter} \times \SI{219}{\milli\meter}$, for DDI contains the interior $\mathcal B_0$ with the two ellipsoidal holes as well as the boundaries of prescribed and unknown nodal forces.}
        \label{fig:geometry_realistic_setup}
    \end{figure}

    To mimic the situation of a real experiment and prepare the data for the dual-stage approach, we perform three steps, which are described below.
    
    \paragraph{FE simulation}
    First, we do a virtual experiment with a \emph{2D plane stress FE simulation} and consider the sample and the applied BCs shown in Fig.~\ref{fig:geometry_realistic_setup}, i.e., the clamping is considered by applying displacement BCs at the two ends of the sample. Linear triangular elements are used.
    Note that only the sample's inner part denoted as $\mathcal B_0$ will be considered later within the DDI.
    Thus, since it is advantageous to have a domain boundary $\partial \mathcal B_0^{\ve f}$ at which the nodal forces ${}^\tau\!\! \ve f^\alpha$ can be specified, cf. Sect.~\ref{sec:data_driven_identification}, the sample is designed such that this cross section is far from the left shoulder and the holes within $\mathcal B_0$. 
    
    As described in Sect.~\ref{sec:dual_stage}, it is only the global testing force ${}^\tau\!\ve F = {}^\tau\! F \ve e_2$ from the load cell of the testing machine that is accessible within each snapshot.
    Thus, to imitate the situation at a testing machine within our numerical experiment, the global testing force is calculated by summing up the nodal forces from the FE simulation at the left clamping.

    To mimic a DIC system, we grab the nodal displacements from the FE simulation.
    For the in-plane components ${}^\tau\!u_1^\alpha$, ${}^\tau\!u_2^\alpha$ this is directly possible.\footnote{We do not take into account here that the displacement at hole edges may not be recorded. For more details on this additional difficulty we refer to \cite{Dalemat2024}.}
    As described in Remark~\ref{remark:compressible}, ${}^\tau\!h^\alpha$  at the nodes can be calculated from the difference in the out-of-plane displacements from top and bottom sides of the sample if we assume to have two DIC systems. 
    In order to simulate this situation, we project the current thickness that is present in the FE calculation at the quadrature points onto the nodes via a weighted average.\footnote{\label{foot:projection}
    The projection of the thickness from the quadrature points to the nodes is done by
    \begin{align*}
        {}^\tau\!h^\alpha = \frac{1}{\bar{S}^\alpha}\sum_{e\in\mathcal E(\alpha)} \!\!\!\!{}^\tau\!h^e S^e \quad \text{with} \quad \bar{S}^\alpha := \!\!\!\!\sum_{e\in\mathcal E(\alpha)} \!\!S^e \; ,
    \end{align*}
    with $\mathcal E(\alpha)$ being the function mapping the node number $\alpha$ to a set containing all element numbers attached to the node, i.e., similar to an inverse coincidence matrix. Note that, since linear triangular elements were used, each element has only one quadrature point, and the quadrature point number $g$ is equal to the element number $e$. Thereby, the area $S^e$ is chosen to be ${}^\tau\!\!J^g_{\!\!\vartriangle}$ or $J^g_{0,\vartriangle}$ depending on the DDI formulation, respectively.
    }
    

    \paragraph{Adding noise}
    In a real experiment, both force measurement and the determination of displacements using DIC are subject to measurement errors. To imitate this, we add artificial noise to both variables, where quantities with noise are marked with a tilde in the following, i.e., $\tilde{(\cdot)}$.

    For the \emph{global testing force}, artificial noise is added for each snapshot by a multiplier based on a continuous uniform distribution, i.e., 
    \begin{align}
        {}^\tau\!\tilde{F} = {}^\tau \! n_{\ve F} {}^\tau\!{F} \quad \text{with} \quad {}^\tau\! n_{\ve F}\sim\mathcal{U}[1-\omega, 1+\omega] \; .    
    \end{align}
    This gives us a signal that is noisy over time, where the noise level is relative to the force magnitude. We choose $\omega = \SI{1e-4}{}$ which corresponds to a load cell with a high accuracy class.

    For the \emph{displacements}, following Flaschel~et~al.~\cite{Flaschel2021}, we assume the same absolute noise level at all nodes and snapshots, independently of the corresponding magnitude of displacement. Therefore, we add artificial noise to the in-plane components and the current thickness, which must be calculated from the displacement difference between the top and bottom of the sample (Remark~\ref{remark:compressible}). However, in the case of \emph{displacement data obtained using DIC}, it can be assumed that DIC software such as ARAMIS smoothes the noisy data so that the noise is not uncorrelated, at least locally in the neighborhood.
    For this reason, a \emph{Gaussian Random Field (GRF)} is used to generate noise within the displacement field instead of random numbers from a normal distribution:
    \begin{align}
        {}^\tau\! \tilde u_a^\alpha &= {}^\tau\! u_a^\alpha + {}^\tau\!u_a^{\alpha,\text{noise}} 
        \quad \text{with} \quad {}^\tau\!u_a^{\alpha,\text{noise}} =  {}^\tau\!f_a(x({}^\tau\!x_1^\alpha),y({}^\tau\!x_2^\alpha)) \eta \Delta x  \quad \text{and}\\
        {}^\tau\! \tilde h^\alpha &= {}^\tau\! h^\alpha + {}^\tau\!h^{\alpha,\text{noise}}
        \quad \text{with} \quad {}^\tau\!h^{\alpha,\text{noise}} = 2 {}^\tau\!f(x({}^\tau\!x_1^\alpha),y({}^\tau\!x_2^\alpha)) \eta \Delta x \label{eq:noise_h}
        \; .
    \end{align}
    In the equations above, $f_a(x,y), f(x,y) \in [-1,1]$ are normalized noise functions on a regular grid $(x,y)$. To investigate the robustness of the dual-stage approach, the noise factor is varied according to $\eta \in \{\SI{1e-3}{}, \,\SI{5e-4}{}, \,\SI{1e-4}{}, \,\SI{5e-5}{}, \,\SI{1e-5}{}\}$. The geometry dimension is given by $\Delta x = \SI{100}{\milli\meter}$.
    Details on the GRF are given in \ref{app:mimic_experiment}. Note that a factor of two is added to the thickness noise in Eq.~\eqref{eq:noise_h}, as both the DIC measurement on the top and bottom of the sample will be subject to noise.

    \begin{figure}
    	    \centering
    	    \includegraphics[width=\textwidth]{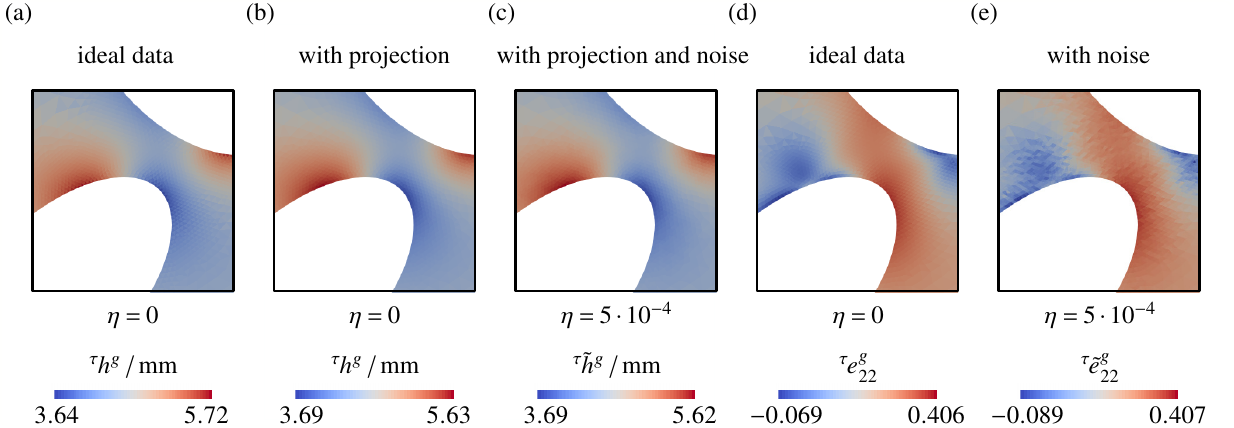}
    	    \caption{Quantitative influence of projecting the thickness from the quadrature points to the nodes and back as well as the noise on DDI field quantities: (a) -- (c) thickness ${}^\tau\! h^g$ as well as (d) -- (e) strain component ${}^\tau\! e_{22}^g$ for updated Lagrangian DDI formulation. Both the projection and the noise have a strong influence on the field data resulting in a change in the ranges of these values.}
    	    \label{fig:nois_projection_error}
    \end{figure}

    \begin{rmk}\label{remark:thickness}
        We would like to point out that the projection and the noise have a particularly strong effect on the thickness as we use absolute noise. The maximum relative error w.r.t. ideal thickness is already $2.3\,\%$ regarding only the projection as well as $2.5\,\%$ including a moderate noise level of $\eta=\SI{5e-4}{}$, cf. Fig.~\ref{fig:nois_projection_error}. This is because the thickness change is very small compared to the in-plane displacements. A plane-strain DDI implementation would therefore be significantly less sensitive but also more complex to realize in experiments. However, the projection leads to a smoothing of the noisy thickness values in comparison to the strain component.
        Actually, the maximum relative error of the noisy strain component ${}^\tau\! \tilde{e}_{22}^g$ w.r.t. the noiseless strain component ${}^\tau\! {e}_{22}^g$ is $17.1\,\%$.
    \end{rmk}
    
    \begin{rmk}\label{remark:denoise}
        Note that it would also be possible to denoise the displacement and force data, e.g., via a suitable filter, as in \cite{Flaschel2021}, just as one would do with data from a real experiment. However, we intentionally do not do this here in order to investigate the robustness of our framework.
    \end{rmk}
    
    \paragraph{Pre-processing the raw data for DDI}

    For the DDI, the nodal forces ${}^\tau \!\! \ve f^\alpha$ on the boundary $\partial \Omega_0^{\ve f}$ are needed and not the global testing force ${}^\tau\!\ve F$. 
    Since perturbations from the holes in the domain and from the clamping are far enough away from the considered cross section, a homogeneous traction ${}^\tau \! \ve p_{\!\ve f} = {}^\tau\!\bar p \ve e_2$ with ${}^\tau\!\bar p = {}^\tau\! F/A_0$ can be assumed at $X_2^\alpha = X_2^\text{min}$, cf. Fig.~\ref{fig:geometry_realistic_setup} as well as Footnote~\ref{foot:anisotropy_constraint}, and we get
    \begin{align}
        {}^\tau\! f_2^\alpha = \sum_{e\in\mathcal E(\alpha)} \ \int\limits_{\partial \Omega_0^{\ve f,e}}
        N^\alpha(X_1,X_2=X_2^\text{min}) {}^\tau\!\bar p h_0 \, \mathrm{d} S  \quad \text{and} \quad 
        {}^\tau\! f_1^\alpha = 0 \; \forall \alpha \text{ with } X_2^\alpha = X_2^\text{min} \;,
    \end{align}
    with $\mathcal E(\alpha)$ being introduced in Footnote~\ref{foot:projection}.
    For the geometry depicted in Fig.~\ref{fig:geometry_realistic_setup}, the error in ${}^\tau\!\bar{\ve p}$ due to the homogeneous traction assumption is less than $1\,\%$. 
    At all other nodes in the domain that do not lie on boundary $\partial\Omega_0^{\ve \zeta}$, ${}^\tau\!\ve f^\alpha= \ve 0$ is specified.

    In addition, the out-of-plane stretch ${}^\tau\!\lambda_3^g={}^\tau\!h^g/h_0$ for determining the ${}^\tau\!e_{33}^g$ or ${}^\tau\!E_{33}^g$ component of the mechanical states is required at the quadrature points. Thus, a projection of the current thickness from the nodes to the quadrature point of each linear triangular element is done via 
    \begin{align}
        {}^\tau\! h^g = \frac{1}{3}\sum_{\alpha \in \mathcal N(g)} \!\!\!{}^\tau\! h^\alpha \; ,   
    \end{align}
    where the set $\mathcal N(g)$ contains all global node numbers $\alpha \in \{1,\dots, N^\text{node}\}$ belonging to element $e$, i.e., similar to a coincidence matrix. Note that even without the noise there is an error due to the projection of the current thickness. However, this would also be unavoidable in a real setup and is therefore deliberately implemented here, cf. Fig.~\ref{fig:nois_projection_error}.

    \subsubsection{Data-driven identification}

    \begin{figure}
    	    \centering
    	    \includegraphics[width=\textwidth]{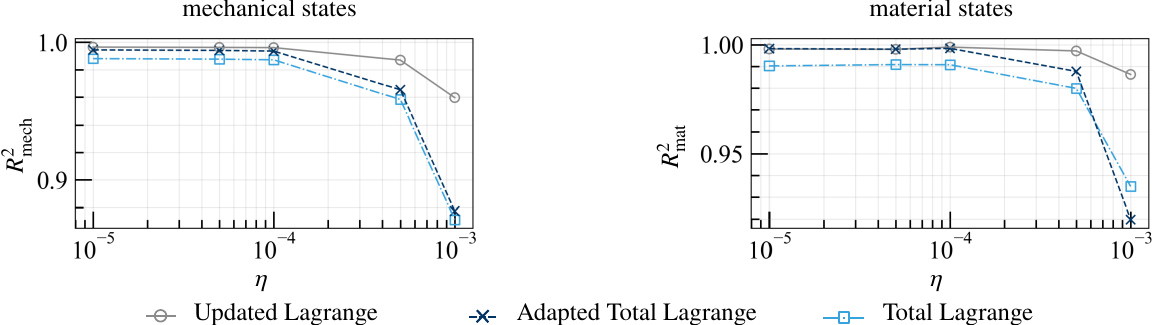}
    	    \caption{Results of the different DDI formulations evaluated with coefficient of determination $R^2$ for mechanical and material states depending on the noise $\eta$ with $N^*/(N^\text{quad}\cdot N^\text{snap}) = \SI{1}{\percent}$ and prescribed pseudo stiffness tensor $\mathbb{C} := \SI{8.3}{\mega\pascal} \,\mathbbm 1$.
    	    }
    	    \label{fig:noisy_data_plot_r2}
    \end{figure}

    As we mimic the situation of a real experiment within this example, the material's stiffness is assumed to be unknown. We must therefore estimate the pseudo stiffness $C$. To this end, we assume an uniaxial stress state with homogeneous field distribution for the sake of simplicity.
    Then, the order of magnitude of the stiffness can be estimated in the sense of 1D linear elasticity for the small strain regime $\sigma = C \varepsilon$. Therefore, we consider the first snapshot $\tau = 1$ to estimate the material's stiffness
    \begin{align}
    C = \frac{\sigma}{\varepsilon}\bigg\vert_{\tau=1}  = \frac{F}{A_0} \frac{l_0}{\Delta l}\bigg\vert_{\tau=1} \approx \frac{\SI{15.096}{\newton}}{500 \,\si{\milli\meter}^2} \frac{219.293\,\si{\milli\meter}}{8.018\,\si{\milli\meter}} \approx 0.82575\,\si{\mega\pascal}\; ,
    \end{align}
    which is equal to the Young's modulus. Obviously, the exact stiffness is underestimated if a uniaxial stress state is assumed for the multiaxial stress state due to the inhomogeneous specimen in this approximation.
    Following the guidelines from Sect.~\ref{sec:assessment_DDI}, the pseudo stiffness tensor is chosen to $\tte C = 10\, C \mathbbm 1 \approx \SI{8.3}{\mega\pascal}\,\mathbbm 1$. For the number of material states we set $N^*/(N^\text{quad}\cdot N^\text{snap}) = \SI{1}{\percent}$.

\begin{figure}
    	    \centering
            \includegraphics[width=\textwidth]{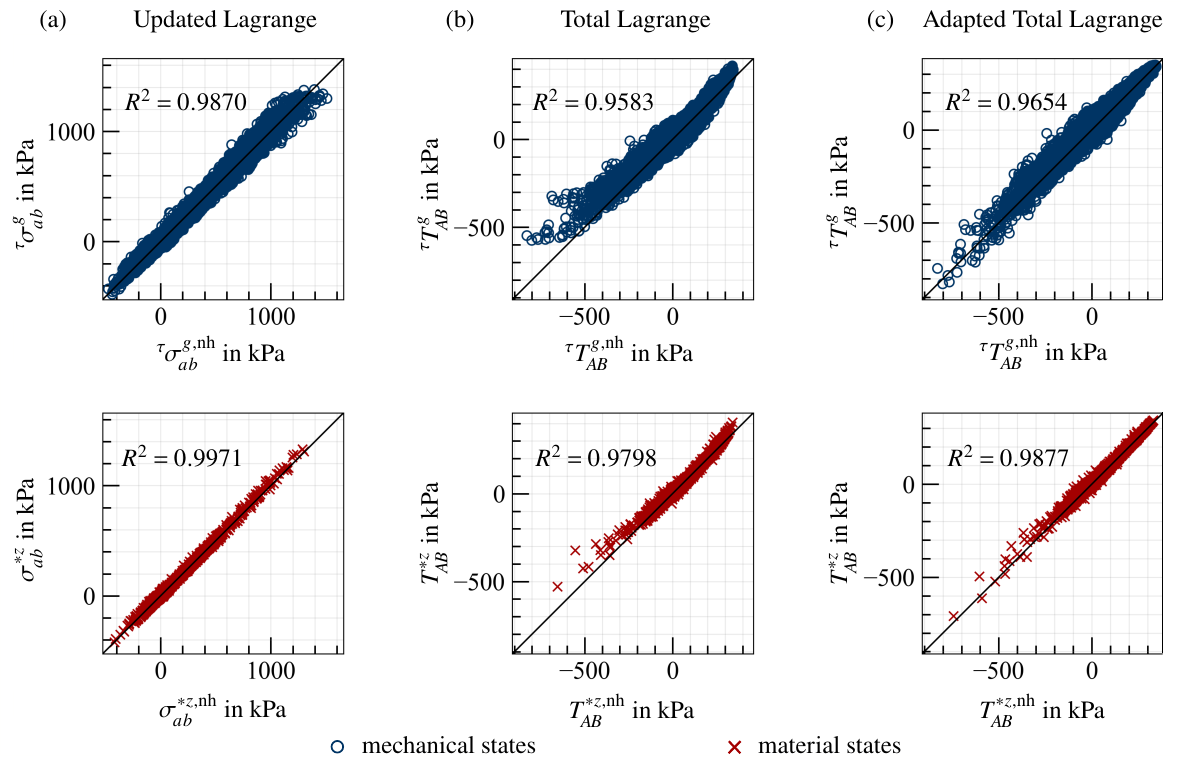}
    	    \caption{Quantitative comparison of Neo-Hooke reference model and DDI results for noise level $\eta=\SI{5e-4}{}$ obtained with (a) Updated-Lagrange-, (b) Total-Lagrange- and (c) Adapted-Total-Lagrange formulation: mechanical states and material states based on noisy data from test specimen discretized with $N^\text{quad} = 7700$ linear elements under loading in each snapshot $\tau \in \{1, \dots, N^\text{snap}\}$ with $N^\text{snap} = 10$ as well as $N^* := \lceil  0.01 \cdot N^\text{quad}\cdot N^\text{snap}\rceil = 770$ as prescribed number of material states of and the given pseudo stiffness tensor $\mathbb{C} := \SI{8.3}{\mega\pascal} \,\mathbbm 1$.
    	    }
    	    \label{fig:noisy_data_plot_correlation}
    \end{figure}

    Using the data from the numerical experiment described in Sect.~\ref{subsec:num_exp} as input, the problem under consideration was solved using all three presented finite strain DDI formulations for all noise factors $\eta$. To evaluate the accuracy for all runs, the $R^2_\text{mech}$ and $R^2_\text{mat}$ values are shown in Fig.~\ref{fig:noisy_data_plot_r2} in dependence of $\eta$, where the ground truth reference is calculated according to Eq.~\eqref{eq:ref_stress} but with the noisy strains, e.g., ${}^\tau\tilde{\te e}^g$. 
    As can be seen, the noise affects the material states less than the mechanical states.
    Although the correlation coefficients $R_\text{mech}^2$ and $R_\text{mat}^2$ are monotonically decreasing with respect to the noise factor, the effect of noise for $\eta \leq 10^{-4}$ is
    negligible and also for $\eta=\SI{5e-4}{}$, the loss of accuracy is acceptable.
    Similar as for the  benchmarks presented in Sect.~\ref{subsec:benchmarks}, the \emph{updated Lagrangian DDI formulation} gives the highest precision followed by the \emph{adapted total Lagrangian DDI formulation} and finally the \emph{original total Lagrangian DDI formulation}. In Fig.~\ref{fig:noisy_data_plot_correlation}, a comparison of all mechanical and material states to the ground truth is shown for $\eta=\SI{5e-4}{}$. 
    As mentioned before, despite the noise, the results are not significantly worse than for the ideal case shown in Fig.~\ref{fig:ideal_data_plot_states}. This is particularly remarkable due to the relatively large error in the thickness and thus the out-of-plane strain, cf. Remark~\ref{remark:thickness}.

    \subsubsection{Physics-augmented neural network}

    \begin{figure}
    	    \centering
    	    \includegraphics[width=\textwidth]{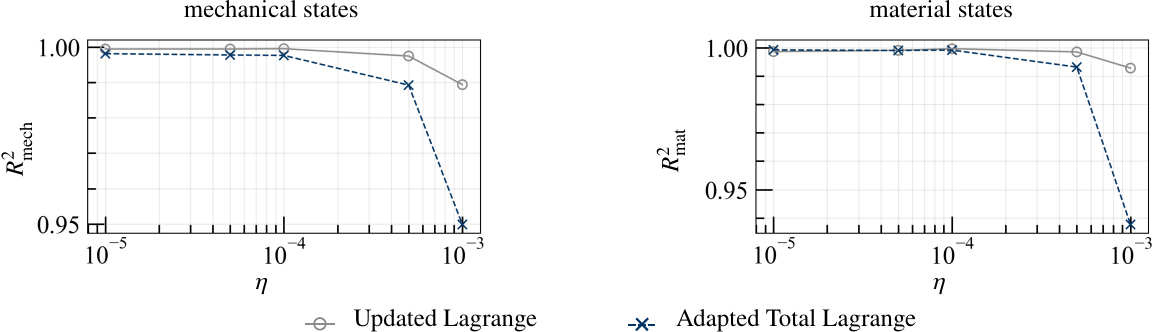}
    	    \caption{Results of the trained PANN with respect to reference data evaluated with coefficient of determination $R^2$ depending on the noise $\eta$. The noisy mechanical and material strains serve as input for the Neo-Hooke reference model.
    	    }
    	    \label{fig:noisy_data_plot_r2_ANN}
    \end{figure}

    \begin{figure}
    	    \centering
            \includegraphics[width=\textwidth]{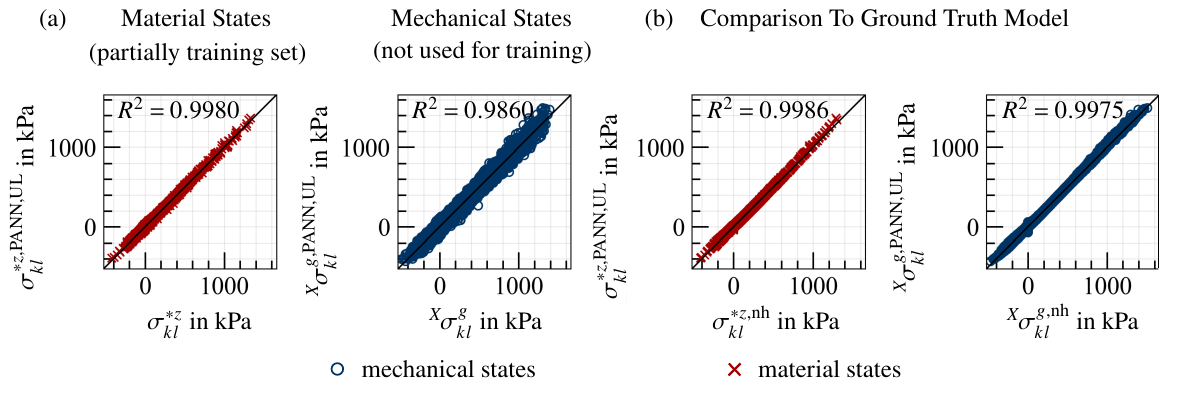}
    	    \caption{
         Quantitative comparison of trained PANN with (a) DDI results of updated Lagrangian formulation and (b) Neo-Hooke reference model for noise level $\eta=\SI{5e-4}{}$ obtained with Updated-Lagrange- formulation: mechanical states and material states based on noisy data from test specimen discretized with $N^\text{quad} = 7700$ linear elements under loading in each snapshot $\tau \in \{1, \dots, N^\text{snap}\}$ with $N^\text{snap} = 10$ as well as $N^* := \lceil  0.01 \cdot N^\text{quad}\cdot N^\text{snap}\rceil = 770$ as prescribed number of material states of and the given pseudo stiffness tensor $\mathbb{C} := \SI{8.3}{\mega\pascal} \,\mathbbm 1$.
    	    }
    	    \label{fig:correlation_noisy_PANN}
    \end{figure}

    In the following, we use the generated data bases, i.e., the material states, to train a PANN model according to Sect.~\ref{sec:neural_networks}, whereby we only use the data generated with the updated Lagrangian DDI formulation and the adapted total Lagrangian DDI formulation. The training is carried out as described in Sect.~\ref{subsec:PANN_training}, whereby we select a ratio of \SI{70}{\percent} \!/\! \SI{30}{\percent} for the division into calibration and test data.

    To evaluate the influence of the noise, the $R^2_\text{mech}$ and $R^2_\text{mat}$ values are shown in Fig.~\ref{fig:noisy_data_plot_r2_ANN} in dependence of $\eta$, where the ground truth reference is again calculated according to Eq.~\eqref{eq:ref_stress} but with the noisy strains, e.g., ${}^\tau \tilde{\te e}^g$. 
    As can be seen, the effect of the noise is negligible for $\eta \leq \SI{1e-4}{}$. A clear influence can be seen for larger values, although this is somewhat smaller than for the DDI itself. This is because a denoising effect occurs due to the PANN, as the loss must be minimized for the noisy database consisting of the material states during training. This means that the ground truth material law is approximated to a good degree by the PANN, even for noisy data, see also \cite{Linden2023}. The $R^2_\text{mech}$ values of the mechanical states are now also higher than with the DDI, which is also due to the fact that the PANN approximates the true material law, which serves as a reference. Even if the strains of the mechanical states are outside the training range, the PANN enables good predictions as it enables a good extrapolation behavior \cite{Linden2023}.

    To better illustrate the denoising effect, a comparison of the PANN predictions for all strains from the mechanical and material states to the ground truth is shown for $\eta=\SI{5e-4}{}$ in Fig.~\ref{fig:correlation_noisy_PANN}.

    \subsubsection{3D Finite Element simulation}

    \begin{figure}
    	    \centering
            \includegraphics[width=\textwidth]{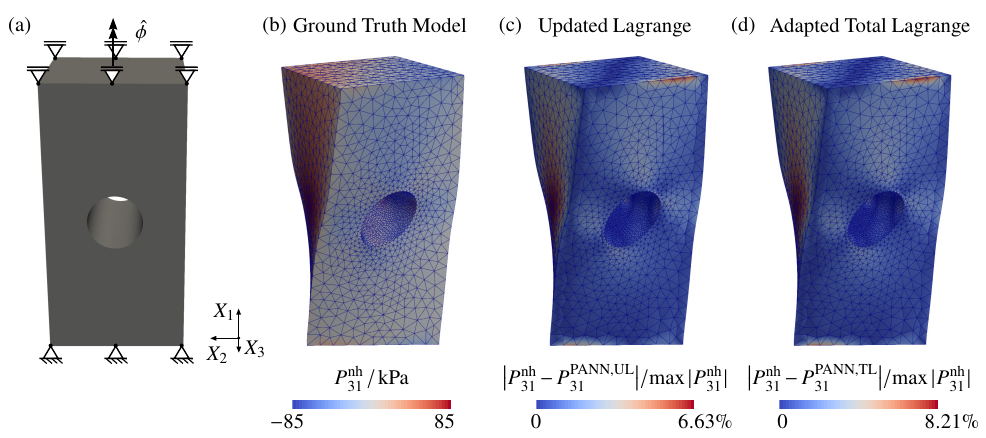}
    	    \caption{FE-simulation of torsional sample: (a) loading and boundary conditions, (b) macroscopic stress field $ P_{31}^\text{nh}$ on the deformed configuration $\mathcal B$ by specifying a distortion of $\hat \phi = \SI{30}{\degree}$, and (c) relative error of the PANN-stress field $P_{31}^{\text{PANN}}$ with respect to $P_{31}^\text{nh}$ trained on the noisy DDI material dataset of the updated Lagrangian formulation as well as (d) of the adapted total Lagrangian formulation. The NNs trained with inhomogenous plane stress data and $N^\text{NN} = 8$ neurons in only one hidden-layer were implemented as constitutive equation each.
   	    }
    	    \label{fig:results_torsion_ellipse_training}
    \end{figure}

    In order to prove the suitability of the developed dual-stage approach, we use the generated PANN models within 3D FE simulations, where we only consider the noise factor $\eta=\SI{5e-4}{}$. We consider the  torsion of a prismatic sample \cite{Kalina2022a,Linden2023}, cf. Fig.~\ref{fig:results_torsion_ellipse_training}(a).
    Both, the ground truth model~\eqref{eq:energy_NH} and the PANN~\eqref{eq:PANN_model}, have been implemented within the FE toolbox FEniCS \cite{Logg2012}.
    The shear stress field $P_{31}^\text{nh}$ generated with neo-Hookean reference model is shown in Fig.~\ref{fig:results_torsion_ellipse_training}(b). Surface plots of relative error measures are given in  Figs.~\ref{fig:results_torsion_ellipse_training}(c) and (d) for the PANN trained with data from the updated Lagrangian DDI and the total Lagrangian DDI, respectively.
    The errors are not negligible but remain within an acceptable range, particularly given the absence of a predefined material law in the DDI framework and the solely reliance on noisy data containing only plane stress states for training of the PANN model. 
    It should be noted again that the maximum relative error of the noisy strain component ${}^\tau\! \tilde{e}_{22}^g$ w.r.t. the noiseless strain component ${}^\tau\! {e}_{22}^g$ is significantly higher with $17.1\,\%$, cf. Fig.~\ref{fig:nois_projection_error}, illustrating the robustness of the framework.    
    The PANN models do not learn the exact material model due to the presence of noise in the data, leading to errors, observed on the the material database, that are of the same order of magnitude in comparison to the reference model.

    Notably, the errors $\max\vert P_{31}^\text{nh} - P_{31}^{\text{PANN,UL}}\vert / \max\vert P_{31}^\text{nh}\vert \approx \SI{1.7}{\percent}$ and $\max\vert P_{31}^\text{nh} - P_{31}^{\text{PANN,TL}}\vert / \max\vert P_{31}^\text{nh}\vert \approx \SI{3.5}{\percent}$ decrease when reducing the noise factor for the displacement field to $\eta = \SI{1e-05}{}$, while maintaining the noise factor for the force field at $\omega = \SI{1e-04}{}$, as in the previous case. This scenario still incorporates the error arising from the projection of thickness values from quadrature points to the nodes and back, cf.~Sect.~\ref{subsec:num_exp}.

    With respect to the ideal dataset, cf.~Sect.~\ref{sec:assessment_DDI}, obtained with $N^*/(N^\text{quad}\cdot N^\text{snap}) = \SI{1}{\percent}$ as well as $C = \SI{1e4}{\kilo\pascal}$, the errors $\max\vert P_{31}^\text{nh} - P_{31}^{\text{PANN,UL}}\vert / \max\vert P_{31}^\text{nh}\vert \approx \SI{0.61}{\percent}$ and $\max\vert P_{31}^\text{nh} - P_{31}^{\text{PANN,TL}}\vert / \max\vert P_{31}^\text{nh}\vert \approx \SI{1.4}{\percent}$ are significantly reduced when considering $\eta = 0$ and $\omega = 0$. Note that no thickness projection from quadrature points to nodes was made in the idealized case.
    
    Evidently, the noise level has a substantial impact on the quality of the material database, which serves as both training and test data for the PANN. In practical applications, denoising algorithms are crucial and would typically be employed to preprocess the input data for the DDI framework. However, in this investigation, the authors deliberately examined the worst-case scenario by utilizing noisy field data without prior denoising, thereby evaluating the robustness of the proposed dual-stage framework.

    The fact that it is possible to carry out 3D simulations with the PANN model at all is due to the chosen formulation using invariants. This is because a large range of 3D states can be covered by plane stress states in the invariant space. The model therefore only has to extrapolate moderately. This has been intensively investigated in \cite{Kalina2022a,Linden2023}.

    \section{Conclusion}\label{sec:conclusion}

    In this work, we present a novel \emph{data-driven dual-stage approach} for the automated constitutive modeling of hyperelasticity, featuring three finite strain versions of the DDI, including two novel total Lagrangian formulations. The overall framework consists of two steps building the core of the approach, which only requires experimentally measurable data. Starting from a (virtual) experiment, full-field measurements are conducted to determine the displacement field and global testing force, with the process referred to as data mining. Then, in the first step of the dual-stage approach, DDI \cite{Leygue2018} is applied to generate a data set consisting of stress and strain tuples. In the third step, the data is used to calibrate a hyperelastic PANN model \cite{Linden2023}. After running the framework, the calibrated PANN can be used in 3D FE simulations.
	The ability of our approach is investigated by applying it to several numerical examples, where a realistic experimental setup is mimicked. Therefore, two-dimensional synthetic data are generated by using a reference constitutive model. To assess the accuracy of our framework, FE simulations with the calibrated PANN are compared to the reference model. 
	
	It turned out that our approach enables to generate accurate constitutive models by only requiring experimentally measurable data as input, i.e., displacement field and global testing force. Also for noisy input data, the deviations to the ground truth model are small even in 3D FE simulations. Our work is a further step in the direction of a paradigm shift in constitutive modeling: from human-based approaches towards fully automated data-driven frameworks. 	
	In summary, the presented data-driven dual-step approach is an efficient methodology to generate hyperelastic models from experiments with full-field measurement. By applying DDI to generate stress-strain data without making constitutive assumptions and subsequently using the data to train a flexible PANN enabling extrapolation afterwards, we combine advantages of two data-driven approaches to build an automated framework.
	
	Various extensions of our approach are planned for the future. For example, the application to anisotropic problems \cite{Fuhg2022b,Kalina2025} is planned. In addition, an extension towards inelasticity, i.e., viscoelasticity \cite{Rosenkranz2024,Abdolazizi2023a,Tac2024a,Holthusen2024} and plasticity \cite{Vlassis2021,Malik2021,Meyer2023a,Fuhg2023,Boes2024} is possible.

	\section*{Acknowledgment}
	All presented computations were performed on a PC-Cluster at the Center for Information Services and High Performance Computing (ZIH) at TU Dresden. The authors thus thank the ZIH for generous allocations of computer time. Furthermore, the authors thank the German Research Foundation (DFG) for the support within the Research Training Group GRK 2868 D${}^3$ -- Project Number 493401063. The authors also thank Alexander Raßloff and Paul Seibert for the help with the implementation of the algorithm for generating the noisy data.
    Finally, the authors would like to thank Franz Hirsch and Philipp Metsch for providing the serverjob scripts to communicate with the HPC-Cluster.
	
	\section*{CRediT authorship contribution statement}
	\textbf{Lennart Linden} and \textbf{Karl A. Kalina} contributed equally to the paper.
    \textbf{Lennart Linden:} Conceptualization, Formal analysis, Investigation, Methodology, Software, Validation, Visualization, Writing - original draft, Writing - review and editing. 
    \textbf{Karl A. Kalina:} Conceptualization, Formal analysis, Methodology, Software, Writing - original draft, Writing - review and editing. 
    \textbf{J\"org Brummund:} Conceptualization, Formal analysis, Methodology, Writing - review and editing.
    \textbf{Brain Riemer:} Software, Writing - review and editing.
    \textbf{Markus K\"astner:} Funding acquisition, Resources, Writing - review and editing.
	
    \appendix

    \section{Details on DDI}\label{ddi_details}
    \subsection{Formulation of the DDI problem}\label{app:formulation_ddi}
    We start with the DDI formulation, which is based on an \emph{updated Lagrangian FE formulation}. To motivate the set of equations, we start by formulating a problem that is continuous in time and space and derive the discrete equations from it.
    
    \subsubsection{Continuous DDI formulation}\label{app:continuous_ddi}
    The primary motivation behind DDI arises from energy considerations in a mechanical system. Based on the elastic stored energy of a continuum, the method aims to minimize the distance between the mechanical states, i.e., field data $\te e(\ve X,t)$, $\teg \tau(\ve X,t)$, and the corresponding material states of the material database, i.e., $\te e^{*z}$, $\teg \tau^{*z}$ with $z\in \{1,\dots,N^*\}$, under the condition that the balance of linear momentum is fulfilled in weak form. Thereby, the material states are allocated to the mechanical states via the mapping 
    $\hat s: \mathcal B_0 \times \mathcal T_\text{DDI} \to \{1, \dots, N^*\}, \; (\ve X,t) \mapsto \hat s(\ve X,t)$ with $\mathcal T_\text{DDI}:= [t_0,t_\text{end}]\subset\R$, cf. Fig.~\ref{fig:ddi_idea}(a) for an illustration of this idea. 
    One of the key aspect of DDI is to use an energetically conjugated pair of strain and stress, here $(\te e, \teg \tau)$. Thus, for the continuous case, this results in the loss function
    \clearpage
    \begin{align}
    \label{eq:continiousUL}
    \begin{split}
        L^\text{UL} &:= \int\limits_{\mathcal T_\text{DDI}} \!\! \int\limits_{\mathcal B_0}  \frac{1}{2} \left(
        \te e(\ve X,t) -  \te e^{*\hat s(\ve X,t)}
        \right) : \ttes t(\ve X,t) : \left(
        \te e(\ve X,t) -  \te e^{*\hat s(\ve X,t)} 
        \right) \, \mathrm d V_0 \, \mathrm d t \\
        &+ \int\limits_{\mathcal T_\text{DDI}} \!\! \int\limits_{\mathcal B_0} \frac{1}{2} \left(
        \teg \tau(\ve X,t) -  \teg \tau^{*\hat s(\ve X,t)}
        \right) : \ttes t^{-1}(\ve X,t) : \left(
        \teg \tau(\ve X,t) -  \teg \tau^{*\hat s(\ve X,t)} 
        \right) \, \mathrm d V_0 \, \mathrm d t \\
        &+ \lambda(t) \Bigg[
        \int\limits_{\mathcal T_\text{DDI}}  \int\limits_{\partial\mathcal B_0^{\ve f}} 
        J(\ve X,t) \ |\te F^{-T}(\ve X,t) \cdot \ve N(\ve X,t)| \ \ve t_{\!\ve f}(\ve X,t) \cdot \ve w(\ve X,t)
        \, \mathrm d A_0 \, \mathrm d t \\
        &+  \int\limits_{\mathcal T_\text{DDI}} \int\limits_{\partial\mathcal B_0^{\ve \zeta}} 
        J(\ve X,t) \ |\te F^{-T}(\ve X,t) \cdot \ve N(\ve X,t)| \ \ve t_{\ve \zeta}(\ve X,t) \cdot \ve w(\ve X,t)
        \, \mathrm d A_0 \, \mathrm d t \\
        &-  \int\limits_{\mathcal T_\text{DDI}} \!\! \int\limits_{\mathcal B_0} 
         \teg \tau(\ve X,t) : \sym(\nabla \ve w(\ve X,t))
        \, \mathrm d V_0 \, \mathrm d t
        \Bigg]
    \end{split}    
    \end{align}
    for the \emph{updated Lagrangian DDI formulation}, where $\ve w(\ve X,t)\in\Lspace{1}$ is a weighting function, $\lambda(t) \in \R$ is a Lagrange multiplier and $\ttes t \in \Lspace{4}$ is a pseudo stiffness. Although the first two integrands of Eq.~\eqref{eq:continiousUL} take the form and units of an energy density, it does not correspond to any actual energy within the mechanical system, cf.~\cite{Dalemat2024}. Instead, the pseudo stiffness tensor $\ttes t$ serves to balance and weight the contributions of strains and stresses.
    In addition, the last three summands of Eq.~\eqref{eq:continiousUL} represent the balance of linear momentum in weak form,  incorporated via the Lagrange multiplier. Thereby, $\partial \mathcal B_0^{\ve f}$ and $\partial \mathcal B_0^{\ve \zeta}$ denote boundaries with prescribed traction $\ve t_{\!\ve f}(\ve X,t) \in \Lspace{1}$ and unknown traction $\ve t_{\ve \zeta}(\ve X,t) \in \Lspace{1}$ on the domain boundary $\partial \mathcal B_0 = \partial \mathcal B_0^{\ve f} \cup \partial \mathcal B_0^{\ve \zeta}$, respectively. Thereby, the prescribed traction $\ve t_{\!\ve f}(\ve X,t) \in \Lspace{1}$ may be zero on parts of the boundary $\mathcal B_0^{\ve f}$.

    \subsubsection{Discrete DDI formulation}\label{app:discrete_ddi}
    Of course, the continuous problem given by Eq.~\eqref{eq:continiousUL} cannot be solved in the specified form. Thus it is not possible to determine a displacement field in an experiment but only displacements at a finite number of control points. Similarly, the displacement field can only be determined for a finite number of time instants, referred to as snapshots in the following.
    To derive the discrete form we take several steps which are described in the following.
    
    First, to arrive at a typical updated Lagrangian formulation, we use the relations $\teg \tau = J \teg \sigma$, $\ttes t = J \ttes c$ and $\mathrm d V = J \mathrm d V_0$, $\mathrm d A = J |\te F^{-T} \cdot \ve N|\mathrm d A_0$, which results in a transformation of the integration domain to the current configuration and to
    $\teg \tau$ being replaced by $\teg \sigma$ as well as $\ttes t$ by $\ttes c$.
    
    Then, we replace the time integral with a sum over snapshots $\tau\in\{1,2,\ldots,N^\text{snap}\}$ and restrict ourselves to a thin sample with reference thickness $h_0\in\R_{>0}$, which allows the \emph{plane stress assumption}, and approximate the geometry with a 2D FE mesh consisting of triangles with domain $\Omega_0^e$ in the undeformed state and ${}^\tau\!\Omega^e$ at snapshot $\tau$, i.e., 
    \begin{align}\label{eq:finite_elements}
        \mathcal B_0 \approx \bigcup_e \Omega_0^e  h_0 \quad \text{and} \quad
        \mathcal B \approx \bigcup_e
        \int\limits_{{}^\tau\!\Omega^e} \frac{{}^\tau\! h({}^\tau \!x_1, {}^\tau\! x_2)}{A_e} \, \mathrm{d} A {}^\tau\!\Omega^e\, ,
    \end{align}
    where ${}^\tau\! h({}^\tau \!x_1, {}^\tau\! x_2)\in \R_{>0}$ is the location-dependent deformed thickness at snapshot $\tau$ weighted with corresponding element area $A_e$.
    Thereby, ${}^\tau \!x_1, {}^\tau\! x_2\in \R$ are the in-plane coordinates of the position vector $\ve x = \ve \varphi(\ve X, \tau) = \ve X + \ve u(\ve X,\tau) \in {}^\tau\!\Omega^e$ in element $e \in \{1,\dots,N^\text{elem}\}$.
    
    In addition, we approximate the in-plane displacement at snapshot $\tau$ within each element $e$ via shapefunctions:
    \begin{align}
    {}^\tau\! u_a({}^\tau\! x_1, {}^\tau\! x_2) \approx \sum_{\alpha \in \mathcal N(e)} N^\alpha({}^\tau\! x_1, {}^\tau\! x_2) {}^\tau\! u_a^\alpha\, ,
    \end{align}
    where again the set $\mathcal N(e)$ contains all global node numbers $\alpha\in\{1,2,\ldots,N^\text{node}\}$ belonging to element $e$ and ${}^\tau\! u_a^\alpha$ being the nodal displacements with the index $a\in\{1,2\}$.  
    Similarly, we approximate ${}^\tau\!\lambda{}^\tau\! w_a({}^\tau\! x_1,{}^\tau\! x_2)$ at snapshot $\tau$ by 
    \begin{align}
        {}^\tau\!\lambda{}^\tau\! w_a({}^\tau\! x_1, {}^\tau\! x_2) \approx \sum_{\substack{\alpha \\ \pi(\alpha) = 1}} N^\alpha({}^\tau\! x_1, {}^\tau\! x_2) {}^\tau\! 
        \eta_a^\alpha + \sum_{\substack{\beta \\ \pi(\beta) = 0}} N^\beta({}^\tau\! x_1, {}^\tau\! x_2) {}^\tau\!\mu_a^\beta \; ,
    \end{align}
    where the function $\pi(\alpha)$ is defined in Eq.~\eqref{eq:pi} and the nodal values ${}^\tau\! \eta_a^\alpha$ as well as ${}^\tau\!\mu_a^\alpha$ take the role of Lagrange multipliers.
    This is to distinguish between nodes lying on $\partial \Omega_0^{\ve \zeta}$, i.e., the traction $\ve t_{\ve \zeta}$ is not prescribed and thus unknown, and nodes lying on $\partial \Omega_0^{\ve f}$, i.e., the traction $\ve t_{\!\ve f}$ is prescribed, or nodes lying in interior $\Omega_0$. Thus, we get the external nodal forces
    \begin{align}
        {}^\tau\!\! \zeta_a^\beta &= \sum_{e\in\mathcal E(\beta)} \ \int\limits_{\partial {}^\tau\!\Omega^{\ve \zeta,e}}
        N^\beta({}^\tau\! x_1, {}^\tau\! x_2) \ t^{\ve \zeta}_a({}^\tau\! x_1, {}^\tau\! x_2) \ {}^\tau\!h({}^\tau\! x_1, {}^\tau\! x_2)\, \mathrm{d} S \, , 
        \\
        {}^\tau\!\! f_a^\alpha &= \sum_{e\in\mathcal E(\alpha)} \ \int\limits_{\partial {}^\tau\!\Omega^{\ve f,e}}
        N^\alpha({}^\tau\! x_1, {}^\tau\! x_2) \ t^{\ve f}_a({}^\tau\! x_1, {}^\tau\! x_2) \ {}^\tau\!h({}^\tau\! x_1, {}^\tau\! x_2) \, \mathrm{d} S \, ,
    \end{align}
    with $\mathcal E(\alpha)$ being introduced in Footnote~\ref{foot:projection}, as well as ${}^\tau\!\! f_a^\alpha = 0$ for all $(X_1^\alpha, X_2^\alpha)\not\in \partial \Omega^{\ve \zeta}_0 \cup \partial \Omega^{\ve f}_0$ with $X_1^\alpha, X_2^\alpha \in \R$ denoting the in-plane nodal coordinates in the undeformed state $\Omega_0$. The in-plane components of the expression ${}^\tau\!\lambda \sym(\nabla {}^\tau\!\ve w)$ are calculated by
    \begin{align}
        \sym\left({}^\tau\!\lambda{}^\tau\! w_{a,b}({}^\tau\! x_1, {}^\tau\! x_2)\right) \approx \sym\Bigg( \sum_{\substack{\alpha \\ \pi(\alpha) = 1}} N^\alpha_{,b}({}^\tau\! x_1, {}^\tau\! x_2) {}^\tau\! 
        \eta_a^\alpha +  \sum_{\substack{\beta \\ \pi(\beta) = 0}} N^\beta_{,b}({}^\tau\! x_1, {}^\tau\! x_2) {}^\tau\!\mu_a^\beta\Bigg) \; , \label{eq:w_grad}
    \end{align}
    where $b$ matrices as common in FE codes are used to calculate the symmetric part given in Eq.~\eqref{eq:w_grad}, e.g., we get
    \begin{align}
    \sym\Bigg( \sum_{\substack{\alpha \\ \pi(\alpha) = 1}} N^\alpha_{,b}({}^\tau\! x_1, {}^\tau\! x_2) {}^\tau\! \eta_a^\alpha\Bigg) = \sum_{\substack{\alpha \\ \pi(\alpha) = 1}} \frac{1}{2}\left(N^\alpha_{,b}\delta_{ac} + N^\alpha_{,a}\delta_{bc}\right) \! ({}^\tau\! x_1, {}^\tau\! x_2) \ {}^\tau\!\eta_c^\alpha = \sum_{\substack{\alpha \\ \pi(\alpha) = 1}} b_{abc}^\alpha({}^\tau\! x_1, {}^\tau\! x_2) {}^\tau\! \eta_c^\alpha\,.
     \end{align}
    
    To solve the remaining integrals in space, we use the isoparametric concept for the description of the geometry, transform the integration domain from physical coordinates $({}^\tau\! x_1, {}^\tau\! x_2)$ to natural coordinates $(\xi_1,\xi_2)$ and
    use Gaussian quadrature with only one quadrature point, e.g., for the integration of a scalar field ${}^\tau T({}^\tau\! x_1, {}^\tau\! x_2)$ over the element volume it follows
    \begin{align}
        \int\limits_{{}^\tau\!\Omega^e} {}^\tau T({}^\tau\! x_1, {}^\tau\! x_2) {}^\tau\!h({}^\tau\! x_1, {}^\tau\! x_2) \, \mathrm d A = \int\limits_{\Omega_{\vartriangle}} \vert{}^\tau\!J_{\!\!\vartriangle}^e(\xi_1,\xi_2)\vert \ {}^\tau T(\xi_1,\xi_2) {}^\tau\! h(\xi_1,\xi_2) \, \mathrm d A_{\!\vartriangle} \approx w^g {}^\tau\!J^g_{\!\!\vartriangle} {}^\tau T^g h^g\; ,
    \end{align}
    with ${}^\tau\!J^g_{\!\!\vartriangle}\in\R_{>0}$ being the determinant of the Jacobian matrix regarding to the coordinate transformation and $w^g\in \R$ the quadrature weight. The superscript index $g$ means that a variable is evaluated at the single quadrature point of the linear element $e$, i.e., the quadrature point number $g$ is equal to the element number $e$, expressed as $g=e$.
    Thus, since the mechanical states only have to be calculated at the quadrature points with in-plane coordinates $({}^\tau\! x_1^g, {}^\tau\! x_2^g)$ at each snapshot, we replace the continuous mapping $\hat s(\ve X,t)$ by the discrete mapping $s: \{1, \dots, N^\text{quad}\} \times \{1, \dots, N^\text{snap}\} \to \{1, \dots, N^*\}, \; (g,\tau) \mapsto s(g,\tau)$ in order to link the quadrature point $g\in\{1,2,\ldots,N^\text{quad}\}$ at snapshot $\tau$ to an entry in the material database. 
    The in-plane components ${}^\tau\! e_{ab}^g$ of the mechanical strain states can be determined from the in-plane coordinates of the deformation gradient at quadrature point $g$, which is calculated from the nodal displacements via 
    \begin{align}
    {}^\tau\!F^g_{ab} = \delta_{ab} +  \sum_{\alpha \in \mathcal N(g)} N^{\alpha g}_{,b} {}^\tau\! u^\alpha_a\,.
    \end{align}
    The out-of-plane component ${}^\tau\! e_{33}^g$ can be determined from the stretch ${}^\tau\! \lambda_3^g = {}^\tau\! h^g/h_0$, cf. Remark~\ref{remark:compressible}.
    In summary, the preceding derivations ultimately lead to the \emph{discrete loss function}~\eqref{eq:snapshots_loss_updated_lagrange}, which forms the basis for the decoupled algorithm introduced in \ref{app:updated_ddi}.
    \begin{rmk}\label{remark:anisotropic}
        Note the following for \emph{anisotropic materials}. Constitutive models are typically formulated using invariants of the right Cauchy-Green deformation tensor $\te C$ and a set of structure tensors, e.g., for transverse isotropy the set is $\mathcal S:=\{\te G\}$ with $\te G \in \Sym$ \cite{Kalina2025}. Since the missing information about the orthogonal tensor $\te R^{*z} \in \SO(3)$ characterizing the rigid body rotation regarding the deformations of the material states, it is not possible to calculate $\te b^{*z}$ from $\te C^{*z}$.
        Therefore, either a total Lagrangian DDI formulation yielding $\mathcal T^{*z} = (\te E^{*z}, \te T^{*z})$ or an interpolation-based adaptation for the updated Lagrangian DDI formulation is required. In contrast to the mechanical states, no unique deformation gradient exists for a given material state. The authors propose assigning the closest admissible deformation gradient $\te F^{*z} \in \GL^+(3)$ to a material state based on those of the mechanical states, assuming that similar strain states correspond to similar deformation states. 
        This can be achieved by the common Log-Euclidean mean approach. Assume that there exist $\te e_1, \dots, \te e_n \in \Sym$ of the mechanical states that lie in the neighborhood of $\te e^{*z}$ with respect to a given norm and tolerance. For the corresponding $\te F_1, \dots, \te F_n \in \GL^+(3)$, we can exploit the Lie group structure to define a weighted mean $\te F_\text{mean} = \exp(\sum_{i=1}^n w_i \ln(\te F_i))$, that preserves the property of having a positive determinant. Thereby, the weights $w_i$ satisfy $\sum_{i=1}^n w_i = 1$, and $\ln(\cdot)$ and $\exp(\cdot)$ denote the tensor logarithm and exponential, respectively. Since for any $\te A \in \Lspace{2}$ it holds that $\det(\exp(\te A)) = \exp(\trace(\te A)) > 0$, the resulting $\te F_\text{mean}$ automatically belongs to $\GL^+(3)$.
        Note that this method requires the $\te F_i$ to be sufficiently close so that the principal branch of the tensor logarithm can be used reliably. Alternative approaches, such as computing the Karcher mean on the Lie group, are available but typically involve higher computational complexity.
    \end{rmk}

    \subsection{Updated Lagrangian DDI algorithm}\label{app:updated_ddi}
    The algorithm for the updated Lagrangian formulation of the data-driven identification is shown in Alg.~\ref{alg:data-driven_updated_Lagrange}.
    The interested reader is referred to Sect.~\ref{sec:ddi_updated_lagrange} for a detailed discussion of the underlying equations and their derivation, which are essential for the implementation.

    \begin{rmk}\label{remark:algorithm}
        Within this remark we give some useful hints for the decoupled algorithm.

        First, the discrete mapping $s(g,\tau)$ must be initialized in step (I) of the Alg.~\ref{alg:data-driven_updated_Lagrange}, whereby the material states are initially unknown. Thus, a random assignment is possible or a clustering based on the strains of the mechanical states can be chosen. 
        As a robust approach for this, k-means clustering is selected, where $N^*$ clusters are selected based on the strains of the mechanical states~\cite{Leygue2018}.
        After the first iteration of the DDI algorithm, the material states -- comprising both stresses and strains -- are already available. However, since no constitutive model is used in this framework, it remains unclear whether a given strain state corresponds to a tensile or compressive stress state. To address this ambiguity, it may be beneficial to perform a one-time reinitialization of the mapping.
        The authors therefore suggest that this reinitialization should rely solely on the stresses of the material states, ensuring that the range of stress values in the material states is better represented and aligned with the expected physical behavior. This step will lead to appropriate clusters of tensile and compressive stress states and can also enhance the convergence properties of subsequent iterations.

        Then, step (II) of the Alg.~\ref{alg:data-driven_updated_Lagrange} requires solving a large linear equation system
    \begin{align}
    \label{eq:equation_system}
        \begin{bmatrix}
            {}^1\M{k} &           &   &   & {}^1\M{s}^{(i)} \\
                    & {}^2\M{k}   &   &   & {}^2\M{s}^{(i)} \\
                    &           & \ddots &  & \vdots \\
                    &           &   & {}^{N^\text{snap}}\!\M{k} & {}^{N^\text{snap}}\!\M{s}^{(i)} \\[6pt]
            {}^1\M{s}^{(i)T} & {}^2\M{s}^{(i)T} & \dots & {}^{N^\text{snap}}\!\M{s}^{(i)T} & \M{0}   
        \end{bmatrix}
        \cdot
        \begin{bmatrix}
            {}^1\Vg{\eta}^{(i)} \\ {}^2\Vg{\eta}^{(i)} \\ \vdots \\ {}^{N^\text{snap}}\!\Vg{\eta}^{(i)} \\[6pt] \Vg{\sigma}^{*(i)}
        \end{bmatrix} =
        \begin{bmatrix}
            {}^1\V{f} \\ {}^2\V{f} \\ \vdots \\ {}^{N^\text{snap}}\!\V{f} \\[6pt] \V{0}
        \end{bmatrix}\; ,
    \end{align}
    obtained from Eqs.~\eqref{eq:snapshots_equation_UL_system_1} and \eqref{eq:snapshots_equation_UL_system_2} with the dimensionalities of the relevant quantities being
    \begin{align}
    {}^\tau\!\M{k} \in \R^{M \times M}, \ {}^\tau\!\M{s}^{(i)} \in \R^{M \times 3 N^*}, \ {}^\tau\Vg{\eta}^{(i)} \in \R^{M}, \ \Vg{\sigma}^{*(i)} \in \R^{3 N^*}, \ {}^\tau\V{f} \in \R^{M}\; , 
    \end{align}
    where $ M$ represents the number of active nodes, defined as $M := 2 \, \big\vert\{\alpha \in \{1, \dots, N^\text{node}\} \mid \pi(\alpha) = 1\}\big\vert \leq 2 N^\text{node}$.
    Even for nonlinear material responses, this step requires only the solution of a linear system.
    Note that the Lagrange multipliers can be identified as virtual displacements enforcing equilibrium~\cite{Leygue2018}.
    The diagonal blocks in the equation system~\eqref{eq:equation_system} remain constant and the off-diagonal blocks must be updated at each iteration step $i \in \N$ of the DDI algorithm. The system matrix is ill-conditioned, making the problem challenging to solve directly and necessitating the use of iterative solvers for practical stability.
    Due to its sparse structure, iterative methods such as the MINRES method are particularly suitable, leveraging the symmetry and indefiniteness of the matrix to ensure stability and efficiency.
    In addition, please note the rank drop of the stiffness matrices ${}^\tau\!\M{k}$, i.e., $\rank({}^\tau\!\M{k}) = 2 N^\text{nodes} - 3$ holds, if and only if the discrete mapping $\pi$ is not surjective, e.g. $\pi \equiv 1$. This is due to the fact that in the two-dimensional case exactly three rigid body motions are not prevented, resulting in three equations in Eq.~\eqref{eq:snapshots_variation_UL_eta} being trivially satisfied.
    In this case, the system of equations must be partitioned or rows and columns must be eliminated in order to fix the drop in rank. However, this case does not occur if the nodal forces ${}^\tau\!\ve\zeta^\beta$ are present at at least two different nodes in all snapshots.
    In general, the discrete mapping $s(g,\tau)$ is not injective, since one material state can be assigned to multiple quadrature points at the same time. But in most cases, the mapping $s(g,\tau)$ is surjective, i.e., each material state is assigned to any quadrature point at least once. Otherwise, the equation system needs to be modified by row and column elimination for example, since the off-diagonal blocks show a rank drop. This means that an unassigned material state will either not be updated or will be completely eliminated from the list of material states, i.e., $N^*_\text{mod} < N^*$ holds after elimination.

    In contrast, steps (III) and (IV) of the Alg.~\ref{alg:data-driven_updated_Lagrange} are computationally inexpensive, involving straightforward computations such as weighted averages and matrix-vector multiplications in order to calculate the mechanical stresses and material strains.

    In step (V) of the Alg.~\ref{alg:data-driven_updated_Lagrange}, the discrete mapping $s(g,\tau)$ must be updated by assigning to each quadrature point $g$ at snapshot $\tau$ the material state that is closest in terms of the loss function~\eqref{eq:snapshots_loss_updated_lagrange}. This process is computationally expensive, as it requires calculating the distance from each mechanical state to all material states stored in the material database. To improve the efficiency of this step, high-dimensional extensions of algorithms like quad-trees or neighbor-list methods, could be considered. The idea is to focus on spatially neighboring points rather than considering all points. This reduces the computational cost by only evaluating interactions between points that are close in the space of interest, as suggested by~\cite{Leygue2018}.
    
    Finally, the test for convergence can be easily done, in step (VI) of the Alg.~\ref{alg:data-driven_updated_Lagrange}, by comparing the updated mapping $s(g,\tau)$ with the previous one.

    The solution procedure of the DDI depends only on two parameters: the number $N^*$ of material states to be identified and the pseudo stiffness tensor $\ttes c$. 
    Some guide values for those can be found in Sect.~\ref{sec:assessment_DDI} based on a hyperparameter study. It should be noted that this stiffness does not necessarily have to be the same for all quadrature points and can therefore be different in each element as well as in each snapshot, cf.~Sect.~\ref{sec:ddi_total_lagrange} with the adapted total Lagrangian DDI formulation. This can increase flexibility, as the individual errors between mechanical and the associated material states are weighted differently in each element.

    In conclusion, the success of the proposed method relies on a key factor as outlined in~\cite{Leygue2018}: the internal richness of individual data items, characterized by the range of ${}^\tau\te e^{g}$ for a given snapshot $\tau$. This internal diversity inherently couples distinct material and mechanical states through the principle of mechanical equilibrium, thereby ensuring that the material behavior can be identified over a broad range of inhomogeneous\footnote{This is also necessary with regard to the use of a PANN model as a constitutive model, particularly if it is to be trained on the basis of these determined material states, cf.~Sect.~\ref{sec:neural_networks}.} material strains $\te e^{*z}$.
    
    \end{rmk}
    
    %
    \SetKwComment{Comment}{/* }{ */}
    \SetKwInOut{Input}{Require}
    \SetKwInOut{Output}{Result}
    \RestyleAlgo{ruled}
    \DontPrintSemicolon
    \begin{table}
    \small
    \begin{algorithm}[H]
    \label{alg:data-driven_updated_Lagrange}
    \caption{Plane-stress data-driven identification solver for the updated Lagrangian formulation}
    \BlankLine
    \BlankLine
    \Input{
    \textbullet ~ discretized  geometry under loading in the snapshots $\tau \in \{1, \dots, N^\text{snap}\}$ with numbered nodes
    \\
    \phantom{\textbullet ~} $\alpha \in \{1, \dots, N^\text{node}\}$ and numbered quadrature points $g \in \{1, \dots, N^\text{quad}\}$ of the linear elements
    \\
    \ \textbullet ~ connectivity encoded by ${}^\tau\!b_{\!abc}^{\alpha g}$
    \\
    \ \textbullet ~ nodal displacements ${}^\tau\!u_a^\alpha$ and nodal forces ${}^\tau\!\!f_a^\alpha$ with $\pi(\alpha) = 1$
    \\
    \ \textbullet ~ element area ${}^\tau\!\!J^g_{\!\!\vartriangle}$ and thickness ${}^\tau\!h^g$ in deformed configuration 
    \\
    \ \textbullet ~ number of material states $N^*$ and pseudo stiffness $c_{klmn}$
    }
    
    \BlankLine
    \Output{
    \textbullet ~ mechanical stresses ${}^\tau\!\sigma_{ab}^g$ and Lagrangian parameter ${}^\tau\!\eta_a^\alpha$
    \\
    \ \textbullet ~ database of material strains $e_{kl}^{*z}$ and stresses $\sigma_{ab}^{*z}$
    \\
    \ \textbullet ~ discrete mapping $s(g,\tau)$ between mechanical and material states
    \\
    \ \textbullet ~ nodal forces ${}^\tau\!\zeta_a^\beta$ with $\pi(\beta) = 0$ from equilibrium}

    \hrulefill
    
    \BlankLine
    \BlankLine
    \textbf{(I)} $i = 0$. Calculate mechanical strains ${}^\tau\!e_{kl}^g$ from nodal displacements. Initialize discrete mapping $s(g,\tau)_{{}_{(i)}}$.
    
    
    \BlankLine
    \textbf{(II)} Solve the linear equation system for all $\sigma_{{ab}_{(i)}}^{*z}$ and ${}^\tau\!\eta_{c_{(i)}}^\alpha$:
    \begin{align}
    \label{eq:alg_snapshots_equation_UL_system_1}
    &\qquad - \sum_\tau\!\!\!\sum_{\substack{\alpha \\ \pi(\alpha) = 1}}\!\!\!\sum_g w^g {}^\tau\!\!J^g_{\!\!\vartriangle} {}^\tau\!h^g {}^\tau\!b_{\!abc}^{\alpha g} \ \delta_{s(g,\tau)_{{}_{(i)}}\!z} \ {}^\tau\!\eta_{{c}_{(i)}}^\alpha = 0\; ,
       \\ \label{eq:alg_snapshots_equation_UL_system_2}
       &\qquad \!\!\!\!\sum_{\substack{\beta \\ \pi(\beta) = 1}}\!\!\!\! \sum_g w^g {}^\tau\!\!J^g_{\!\!\vartriangle} {}^\tau\!h^g {}^\tau\!b_{\!abc}^{\alpha g} c_{abde} {}^\tau\!b_{\!de\!f}^{\beta g} {}^\tau\!\eta_{{\!f}_{\!(i)}}^\beta + \sum_g w^g {}^\tau\!\!J^g_{\!\!\vartriangle} {}^\tau\!h^g {}^\tau\!b_{\!abc}^{\alpha g} \ \delta_{s(g,\tau)_{{}_{(i)}}\!z} \ \sigma_{{ab}_{(i)}}^{*z} = {}^\tau\!\!f_c^\alpha \; .
    \end{align}
    
    \textbf{(III)} Calculate the mechanical stresses.
    \\
    \For{$\tau = 1,\dots, N^\mathrm{snap}$ $\text{\upshape and}$ $g = 1,\dots, N^\mathrm{quad}$}{
    \begin{align}
    \label{eq:alg_snapshots_equation_UL_update_stress}
    \qquad {}^\tau\!\sigma_{{ab}_{(i)}}^g = \sigma_{{ab}_{(i)}}^{*s(g,\tau)_{{}_{(i)}}} + \!\!\!\!\sum_{\substack{\alpha \\ \pi(\alpha) = 1}}\!\!\!\! c_{abcd} {}^\tau\!b_{\!cde}^{\alpha g} {}^\tau\!\eta_{{e}_{(i)}}^\alpha
    \end{align}
    }

    \BlankLine
    \BlankLine
    \textbf{(IV)} Calculate the material strains.
    \\
    \For{$z = 1,\dots, N^*$}{
    \begin{align}
    \label{eq:alg_snapshots_equation_UL_update_strain}
       \qquad e_{{kl}_{(i)}}^{*z} = \dfrac{1}{\sum\limits_\tau\sum\limits_g w^g {}^\tau\!\!J^g_{\!\!\vartriangle} {}^\tau\!h^g \delta_{s(g,\tau)_{{}_{(i)}}\!z}} \sum_\tau\sum_g w^g {}^\tau\!\!J^g_{\!\!\vartriangle} {}^\tau\!h^g \ \delta_{s(g,\tau)_{{}_{(i)}}\!z} \ {}^\tau\!e_{kl}^g
    \end{align}
    }
 
    \BlankLine
    \BlankLine
    \textbf{(V)} Update discrete mapping $s(g,\tau)_{{}_{(i)}}$.
    \\
    \For{$\tau = 1,\dots, N^\mathrm{snap}$ $\text{\upshape and}$ $g = 1,\dots, N^\mathrm{quad}$}{
    Select $\big(e_{{kl}_{(i+1)}}^{*z}, \sigma_{{ab}_{(i+1)}}^{*z}\big)$ closest to mechanical state $\big({}^\tau\!e_{{kl}_{(i)}}^g, {}^\tau\!\sigma_{{ab}_{(i)}}^g\big)$ with respect to loss function $L^\mathrm{UL}$.
    }
    
    \BlankLine
    \BlankLine
    \textbf{(VI)} Test for convergence.
    \\
    \uIf{$s(g,\tau)_{{}_{(i+1)}} = s(g,\tau)_{{}_{(i)}}$ $\text{\upshape for all}$ $\tau = 1, \dots, N^\mathrm{snap}$ $\text{\upshape and}$ $g = 1, \dots, N^\mathrm{quad}$}
    {
    $\text{\upshape Calculate unknown nodal forces ${}^\tau\!\zeta_a^\beta$ with $\pi(\beta) = 0$ from equilibrium, if requested.}$
    \\
    \textbf{exit}
    }
    \Else{
    $i \gets i + 1$, \textbf{goto} \textbf{(II)}.
    }
    \end{algorithm}
    \end{table}

    \subsection{Adapted total Lagrangian DDI algorithm}\label{app:adapted_lagrange_ddi}
    
    The algorithm for the adapted total Lagrangian formulation of the data-driven identification is shown in Alg.~\ref{alg:data-driven_adapted_total_Lagrange}. Derivations of the corresponding equations are provided in Sect.~\ref{sec:ddi_total_lagrange}.
    The main difference to Alg.~\ref{alg:data-driven_updated_Lagrange} lies in the fact that the pseudo stiffness tensor at each quadrature point is variable, i.e. it depends on the deformation gradient. This leads, among other things, to a modified update step~\eqref{eq:alg_snapshots_equation_TL_update_strain} of the strains of the material states.

    %
    \SetKwComment{Comment}{/* }{ */}
    \SetKwInOut{Input}{Require}
    \SetKwInOut{Output}{Result}
    \RestyleAlgo{ruled}
    \DontPrintSemicolon
    \begin{table}
    \small
    \begin{algorithm}[H]
    \label{alg:data-driven_adapted_total_Lagrange}
    \caption{Plane-stress data-driven identification solver for the adapted total Lagrangian formulation}
    \BlankLine
    \BlankLine
    \Input{
    \textbullet ~ discretized  geometry under loading in the snapshots $\tau \in \{1, \dots, N^\text{snap}\}$ with numbered nodes
    \\
    \phantom{\textbullet ~} $\alpha \in \{1, \dots, N^\text{node}\}$ and numbered quadrature points $g \in \{1, \dots, N^\text{quad}\}$ of the linear elements
    \\
    \ \textbullet ~ connectivity encoded by ${}^\tau\!B_{\!ABc}^{\alpha g}$
    \\
    \ \textbullet ~ nodal displacements ${}^\tau\!u_a^\alpha$ and nodal forces ${}^\tau\!\!f_a^\alpha$ with $\pi(\alpha) = 1$
    \\
    \ \textbullet ~ element area $J^g_{0,\vartriangle}$ and thickness $h^g_{0,\vartriangle}$ in undeformed configuration 
    \\
    \ \textbullet ~ number of material states $N^*$ and pseudo stiffness ${}^\tau C_{K\!L M\!N}^g$
    }
    
    \BlankLine
    \Output{
    \textbullet ~ mechanical stresses ${}^\tau T_{\!AB}^g$ and Lagrangian parameter ${}^\tau\!\eta_a^\alpha$
    \\
    \ \textbullet ~ database of material strains $E_{\!K\!L}^{*z}$ and stresses $T_{\!AB}^{*z}$
    \\
    \ \textbullet ~ discrete mapping $s(g,\tau)$ between mechanical and material states
    \\
    \ \textbullet ~ nodal forces ${}^\tau\!\zeta_a^\beta$ with $\pi(\beta) = 0$ from equilibrium}

    \hrulefill
    
    \BlankLine
    \BlankLine
    \textbf{(I)} $i = 0$. Calculate mechanical strains ${}^\tau\!E_{\!K\!L}^g$ from nodal displacements. Initialize discrete mapping $s(g,\tau)_{{}_{(i)}}$.
    
    
    \BlankLine
    \textbf{(II)} Solve the linear equation system for all $T_{{\!AB}_{(i)}}^{*z}$ and ${}^\tau\!\eta_{c_{(i)}}^\alpha$:
    \begin{align}
    \label{eq:alg_snapshots_equation_TL_system_1}
    &\qquad - \sum_\tau\!\!\!\sum_{\substack{\alpha \\ \pi(\alpha) = 1}}\!\!\!\sum_g w^g J^g_{0,\vartriangle} h^g_{0,\vartriangle} {}^\tau\!B_{\!ABc}^{\alpha g} \ \delta_{s(g,\tau)_{{}_{(i)}}\!z} \ {}^\tau\!\eta_{{c}_{(i)}}^\alpha = 0\; ,
       \\ \label{eq:alg_snapshots_equation_UL_system_2}
       &\qquad \!\!\!\!\sum_{\substack{\beta \\ \pi(\beta) = 1}}\!\!\!\! \sum_g w^g J^g_{0,\vartriangle} h^g_{0,\vartriangle} {}^\tau\!B_{\!ABc}^{\alpha g} {}^\tau C_{\!A\!BD\!E}^g {}^\tau\!B_{\!D\!E\!f}^{\beta g} {}^\tau\!\eta_{{\!f}_{\!(i)}}^\beta + \sum_g w^g J^g_{0,\vartriangle} h^g_{0,\vartriangle} {}^\tau\!B_{\!ABc}^{\alpha g} \ \delta_{s(g,\tau)_{{}_{(i)}}\!z} \ T_{{\!AB}_{(i)}}^{*z} = {}^\tau\!\!f_c^\alpha \; .
    \end{align}
    
    \textbf{(III)} Calculate the mechanical stresses.
    \\
    \For{$\tau = 1,\dots, N^\mathrm{snap}$ $\text{\upshape and}$ $g = 1,\dots, N^\mathrm{quad}$}{
    \begin{align}
    \label{eq:alg_snapshots_equation_TL_update_stress}
    \qquad {}^\tau T_{{\!AB}_{(i)}}^g = T_{{\!AB}_{(i)}}^{*s(g,\tau)_{{}_{(i)}}} + \!\!\!\!\sum_{\substack{\alpha \\ \pi(\alpha) = 1}}\!\!\!\! {}^\tau C_{\!A\!B C\!D}^g {}^\tau\!B_{\!C\!Da}^{\alpha g} {}^\tau\!\eta_{{a}_{(i)}}^\alpha
    \end{align}
    }

    \BlankLine
    \BlankLine
    \textbf{(IV)} Calculate the material strains.
    \\
    \For{$z = 1,\dots, N^*$}{
    \begin{align}
    \label{eq:alg_snapshots_equation_TL_update_strain}
       \qquad E_{{\!K\!L}_{(i)}}^{*z} = \bigg(\sum\limits_\tau\sum\limits_g w^g J^g_{0,\vartriangle} h^g_{0,\vartriangle} {}^\tau C_{K\!L M\!N}^g\delta_{s(g,\tau)_{{}_{(i)}}\!z}\bigg)^{-1} \sum_\tau\sum_g w^g J^g_{0,\vartriangle} h^g_{0,\vartriangle} {}^\tau C_{\!M\!N\!PQ}^g {}^\tau\!E_{PQ}^g \delta_{s(g,\tau)_{{}_{(i)}}\!z}
    \end{align}
    }
 
    \BlankLine
    \BlankLine
    \textbf{(V)} Update discrete mapping $s(g,\tau)_{{}_{(i)}}$.
    \\
    \For{$\tau = 1,\dots, N^\mathrm{snap}$ $\text{\upshape and}$ $g = 1,\dots, N^\mathrm{quad}$}{
    Select $\big(E_{{\!K\!L}_{(i+1)}}^{*z}, T_{{\!AB}_{(i+1)}}^{*z}\big)$ closest to mechanical state $\big({}^\tau E_{{\!K\!L}_{(i)}}^g, {}^\tau T_{{\!AB}_{(i)}}^g\big)$ with respect to loss function $L^\mathrm{TL}$ with adapted pseudo stiffness.
    }
    
    \BlankLine
    \BlankLine
    \textbf{(VI)} Test for convergence.
    \\
    \uIf{$s(g,\tau)_{{}_{(i+1)}} = s(g,\tau)_{{}_{(i)}}$ $\text{\upshape for all}$ $\tau = 1, \dots, N^\mathrm{snap}$ $\text{\upshape and}$ $g = 1, \dots, N^\mathrm{quad}$}
    {
    $\text{\upshape Calculate unknown nodal forces ${}^\tau\!\zeta_a^\beta$ with $\pi(\beta) = 0$ from equilibrium, if requested.}$
    \\
    \textbf{exit}
    }
    \Else{
    $i \gets i + 1$, \textbf{goto} \textbf{(II)}.
    }
    
    \end{algorithm}
    \end{table}
    
    \clearpage
    \section{Gaussian random field}\label{app:mimic_experiment}
    When introducing noise into displacement fields for a two-dimensional geometry, it is crucial to account for correlations between the values at neighboring nodes. Using locally uncorrelated noise, such as Gaussian random variables with no spatial correlation, risks producing unphysical configurations. Specifically, for very small elements, it is possible for three nodes of a linear triangular element to align, causing the element's volume to approach zero. This can lead to numerical instabilities during simulation.

    To mitigate this issue, the noise is modeled as a \emph{Gaussian Random Field} (GRF). 
    The distinguishing characteristic of a GRF is that it is completely  defined by the two-point correlation function~$\mathcal S$, whereas higher-order statistics carry no additional information \cite{Liu2019b}.
    In the present work, the Fourier transform of~$\mathcal S$, also referred to as power spectrum, is modeled for an exponential decay in physical space as
    \begin{align}
       \hat{\mathcal S}(\mathbf{k}) = \exp(-\ell^2 \vert\mathbf{k}\vert^2)\,,
    \end{align}
    where $\hat{(\cdot)}$ indicates Fourier transform, $\vert\mathbf{k}\vert^2$ denotes the squared magnitude of the wavevectors $\mathbf{k} \in \R^2$ on the regular grid and $\ell \in \R_{>0}$ stands for the length scale of the spatial correlation.
    To generate a realization of the random field in real space, a complex field of uncorrelated random variables~$\eta$ is drawn from a normal distribution with zero mean and unit variance. After convolution with the square root of the power spectrum and the inverse Fourier transform performed, this is expressed as
    \begin{align}
       f(x, y) = \mathcal{F}^{-1}\Big[\sqrt{\hat{\mathcal S}(\mathbf{k})} \, \eta(\mathbf{k})  \Big]\,
    \end{align}
    and subsequent normalized.
    Finally, the result is interpolated from the regular grid to the FE mesh by a bilinear interpolation.

    In this paper, the authors choose to generate GRFs on a square regular grid with $N = 4096$ pixels each in $x$- and $y$-direction with decay length $\ell = 1/N$. Afterwards, the regular domain of the GRFs with a rectangular grid $N_x = 1024$ by $N_y = 4096$ is selected.

    \begin{figure}
    	    \centering
    	    \includegraphics[width=\textwidth]{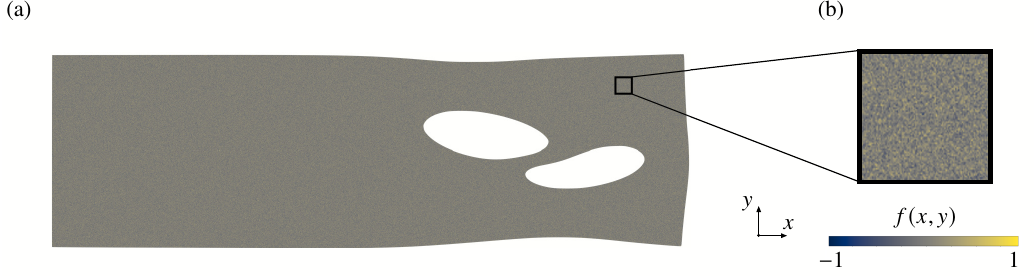}
    	    \caption{Normalized GRF noise on regular grid with $N = 4096$ with decay length $\ell = 1/N$ projected on deformed specimen with $N_x = 1024$ and $N_y = 4096$.
    	    }
    	    \label{fig:grf_noise}
    	\end{figure}


\bibliographystyle{unsrtnat} 
\bibliography{DDI_PANNs.bib}

\end{document}